%% file: CElegans.tex
\documentclass[12pt]{amsart}
\usepackage{setspace}
\usepackage[T1]{fontenc}
\usepackage{tgtermes}
\usepackage{amsmath, amsthm, amsfonts, amssymb}
\usepackage{mathrsfs}
\usepackage[english]{babel}
\usepackage{graphicx}
\usepackage{graphics}
\usepackage{hyperref}
\usepackage{url}
\usepackage[all]{xy}
\usepackage{enumerate}
\usepackage{fancyhdr}
\usepackage{xspace}
\usepackage{verbatim}
\usepackage{epstopdf}
\usepackage{subfloat}
\usepackage[a4paper, landscape]{lscape}
\usepackage{subfigure}

\usepackage{tikz,xcolor,color}
\usetikzlibrary{calc,arrows,decorations.pathmorphing,backgrounds,fit,automata,shapes.geometric,shapes.arrows}
\usetikzlibrary{positioning}
\theoremstyle{definition}

\theoremstyle{remark}

\theoremstyle{plain}

\numberwithin{table}{section}

\newtheorem*{theorem*}{Theorem}

\numberwithin{equation}{section}
\numberwithin{figure}{section}

\newcommand{\round}[1]{{\ooalign{\hfil\raise .10ex\hbox{\scriptsize#1}\hfil\crcr\mathhexbox20D}}}

\long\def\red#1{\textcolor[rgb]{1.00,0.00,0.00}{#1}}
\long\def\red#1{}

\def\s-s{self-similar}

\allowdisplaybreaks[4]
\title[Mathematically Modeling the \emph{C. elegans} Neural Network]{Analyzing properties of the \emph{C. elegans} Neural Network:  Mathematically Modeling a Biological System}

\thanks{Research supported in part by NSF grant DMS-0505622}

\author[D.~J.~Kelleher]{Daniel J.~Kelleher}
\address[D.~J.~Kelleher]{Department of Mathematics, University of Connecticut, Storrs, CT 06269, USA}
\email[D.~J.~Kelleher]{kelleher@math.uconn.edu}
\urladdr{\url{http://www.math.uconn.edu/~kelleher/}}

\author[T.~M.~Reese]{Tyler M.~Reese}
\address[T.~M.~Reese]{Department of Mathematics, University of Connecticut, Storrs, CT 06269, USA}
\email[T.~M.~Reese]{tyler.reese@uconn.edu}

\author[D.~T.~Yott]{Dylan T.~Yott}
\address[D.~T.~Yott]{Department of Mathematics, Boston University, Boston, MA, 02135, USA}
\email[D.~T.~Yott]{dyott@bu.edu}

\author[A.~Brzoska]{Antoni~Brzoska}
\address[A.~Brzoska]{Department of Mathematics, University of Connecticut, Storrs, CT 06269, USA}
\email[A.~Brzoska]{antoni.brzoska@uconn.edu}

\begin{document}
%
%

\begin{abstract}
The brain is one of the most studied and highly complex systems in the biological world.  It is the information center behind all vertebrate and most invertebrate life, and thus has become a major focus in current research.  While many of these studies have concentrated on studying the brain directly, our focus is the structure of the brain itself: at its core an interconnected network of nodes (neurons).  A better understanding of the structural aspects of the brain should elucidate some of its functional properties.  In this paper we analyze the brain of the nematode \emph{Caenorhabditis elegans}.  Consisting of only 302 neurons, it is one of the better-understood neural networks.  Using a Laplacian matrix of the 279-neuron ``giant component" of the network, we use an eigenvalue counting function to look for fractal-like self similarity.  This matrix representation is also used to plot (in eigenfunction coordinates) both 2 and 3 dimensional visualizations of the neural network.  Further analysis examines the small-world properties of the system, including average path length and clustering coefficient.  We then test for localization of eigenfunctions, using graph energy and spacial variance.  To better understand these results, all of these calculations are also performed on random networks, branching trees, and known fractals, as well as fractals which have been ``rewired" to have small-world properties. This analysis is one of many stepping-stones in the research of neural networks.  While many of the structures and functions within the brain are known, understanding how the two interact is also important.  A firmer grasp on the structural properties of the neural network is a key step in this process.
\setcounter{tocdepth}{1}
\tableofcontents
\end{abstract}\maketitle

\section*{Author Summary}
	The brain is the biological center driving all animal life, and certainly our own.  Though fundamentally an interconnected system of nodes, the information-transfer that occurs inside of these networks perpetuates life itself.  As a result, the brain is the focus of immeasurable scientific research.  While many studies clinically analyze the brain and its function, our goal is to look at the brain on a simpler level- as a network of connected points.  Studying the structural connectivity and network properties within the brain will help further our understanding of the organization underlying its many functions.  In this paper, we study the neural network of the nematode worm \emph{Caenorhabditis elegans}.  It is a well-understood system consisting of 302 neurons, making it an excellent candidate for our research. (The human brain is composed of billions of neurons, the connections between which are not completely known).  Through the course of our study, we look for both self-similarity and small-world characteristics in the structure of the \emph{C. elegans} neural network.  A better understanding of the brain's physical configuration will help to reveal the mechanisms behind its structure and function.
\input{introduction}
\input{Results}

\input{Methods2}
\section*{Acknowledgements}
	We thank Dr. Alexander Teplyaev, Department of Mathematics, University of Connecticut, for his guidance in organizing and overseeing our project.  We thank Dr. Dmitri Chklovskii, Howard Hughes Medical Institute, Janiela Farm Research Campus, for allowing us to extend his work and use his adjacency matrices.  We thank Matthew Begu\'e, Department of Mathematics, University of Maryland, for his help with our MATLAB code.  We also thank Lisa Pham, Boston University, for her advice.
	
\section*{Author Contributions}
	DJK conceived and designed the experiments. TMR performed the experiments, analyzed the data, and wrote the paper sections concerning the Eigenvalue Counting Function, Weyl Ratios, and the Eigen Projection Method.  DTY performed the experiments, analyzed the data, and wrote the paper sections concerning Small-World properties.  AB performed the experiments, analyzed the data, and wrote the paper sections concerning Graph Energy and Spacial Variance.  TMR wrote the remainder of the manuscript.  DJK, TMR, AB, and DTY carried out revisions of the manuscript.

\bibliographystyle{plos}
\bibliography{References}
%
\end{document}

%% file: introduction.tex
\section{Introduction}\label{sec:introduction}
	Fractal theory has become an increasingly popular topic of both debate and research in recent years.  Beginning with Mandelbrot's discussion of Britain's immeasurable coastline \cite{MAND1}, fractal theory has found applications in both the mathematics and scientific communities.  In the geometric sense, fractals are objects that exhibit self-symmetry: they exhibit the same pattern on increasingly smaller scales.  In other words, magnifications of smaller portions resemble the whole object.
	
	  More recently, fractal theory has found applications in the biological realm.  Kinetics of ion channels have been modeled with fractal structures \cite{LFKTW2, LLW3}.  Fractal dimension has been used to analyze human EEG signals \cite{PU4} as well as the complex morphology of living cells \cite{BBG5, SLM6}.  The applications of fractal theory in neuroscience have been a particularly prevalent topic of research \cite{FJ7, KHA8, FJ9, W10, M11}.  Glial cells have been analyzed in-depth using fractal dimensions and modeling \cite{SB12, SBLMS13, RSSS14}.  Dendritic branching has been shown to have self-similarity \cite{CSEDHN15, B16}, and other studies have examined fractal patterns in neuron connectivity \cite{S17}.  Further applied research has used three-dimensional fractal structures to approximate the white matter surface of the human brain, based on MRI images \cite{FSCFS18}.
	
	In this paper we use a graph-theoretical approach to probe the structure of the \emph{Caenorhabditis elegans }neural network for self-similar structures.  Advances in graph theory have proven highly useful in analyzing complex neural networks \cite{BO19, SR20, FCR21} as well as underlying motifs in the brain \cite{SK22, IA23}.  These methods have made considerable contributions to our understanding of the structure and function of one of biology's most intricate and important systems.  In this paper we apply these techniques to a physical map of the \emph{C. elegans} brain.  With a well-connected component of only 279 neurons, it is an excellent candidate for graph theoretical research on a complete-brain model.  While \cite{MOOFOK24} presents a geometric structure of this nematode brain, our research continues that of \cite{VPC25} in which Chklovskii et al. propose a finalized schematic of the \emph{C. elegans} neural network.
	
	The \emph{C. elegans} brain is composed of three types of neurons: sensory neurons, motor neurons, and interneurons.  Two types of connections exist between these neurons: chemical synapses and gap junctions.  The gap junction network, which sends electrical signals via ion transport, is an undirected system.  Conversely, chemical synapses possess clear directionality \cite{VPC25}. We are only interested in studying the overall connectivity between neurons.  Thus in order to analyze the structure of the \emph{C. elegans} neural network, we consider only the skeleton of the brain's organization.  Although some neurons share multiple points of contact (they have a multiplicity of connection > 1) we consider this a single connection.  While chemical synapses send directional signals, we only observe that two neurons are connected: regardless of that connection's direction.  As a result, we are able to study a weakly connected network representing only the framework of connections.  (See Methods)
	
	In order to index each of these connections we used the graph Laplacian matrix, $L =[l_{i,j}]$.  For a graph, $G$, we define $d_{v}$ as the degree of a vertex $v$: the number of total connections.  (Note that each vertex represents a neuron).  If vertex $u$ is connected to vertex $w$ then $l_{u,w}=-1$ and $l_{w,u}=-1$.  These correspond to the entries in the $u^{th}$ row and $w^{th}$ column, and the $w^{th}$ row and $u^{th}$ column.  Furthermore, $L_{v,v}=d_{v}$.  All other entries of matrix $L$ are 0.  We also generated similar matrices for randomly generated networks, random branching trees, and known fractals.  
	
	The original goal of this study was to examine the structure of the \emph{C. elegans} neural network for self-similarity.  Through the course of our research we analyzed many other network properties, and compared these results with similar calculations on other systems.  Although fractal theory has repeatedly been applied in neuroscience, in studying the structure of the nematode brain the results are not as simple as saying ``fractal" or ``not-fractal."  Instead we search for \emph{self-similar} structures, as there is no precise definition of what it means for a network to have "fractal properties."  Nevertheless, we uncover several interesting properties of the \emph{C. elegans} neural network.

%% file: Results.tex
\section{Results and Discussion}\label{Results and Discussion}
	Before we could begin our analysis, we had to construct a variety of Laplacian matrices.  For our \emph{C. elegans} model, we derived a Laplacian matrix from the adjacency matrices used in \cite{VPC25} (See Methods).  We already had the Laplacian matrices of known fractals such as the Sierpinski Carpet, the Octagasket, the Hexacarpet, and the Sierpinski Gasket (specific subdivisions were chosen which had similar numbers of vertices as the \emph{C. elegans} neural network).  Next we wrote a series of MATLAB programs which would generate Laplacian matrices with specific conditions.  One produced that of a random network when given a defined number of vertices ($n$) and probability of connection ($p$).  Another produced the Laplacian matrix of a random-branching tree, given the total number of vertices ($n$) as well as the maximum number of branches possible from any point ($m$).  One final program randomly rewired these networks, moving connections from one vertex to another with a given probability, $p$.  Details on these programs can be found in Methods \ref{methodsrandom} and \ref{methods3.8}.

\subsection{The Eigenvalue Counting Function}
	
\begin{figure}[ht]
\centering
\subfigure[\emph{C. elegans} neural network]{
\includegraphics[width=.45\linewidth]{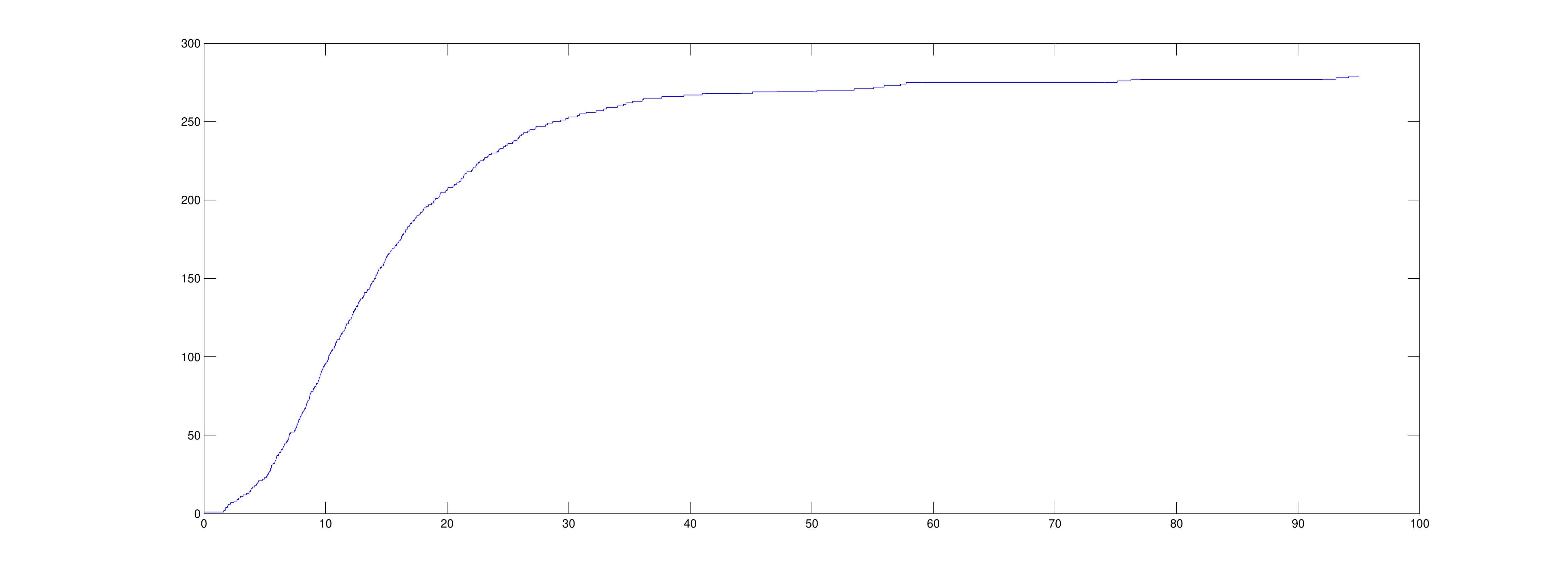}
\label{eigcountworm}}
\subfigure[Sierpinski Gasket, Level 5]{
\includegraphics[width=.45\linewidth]{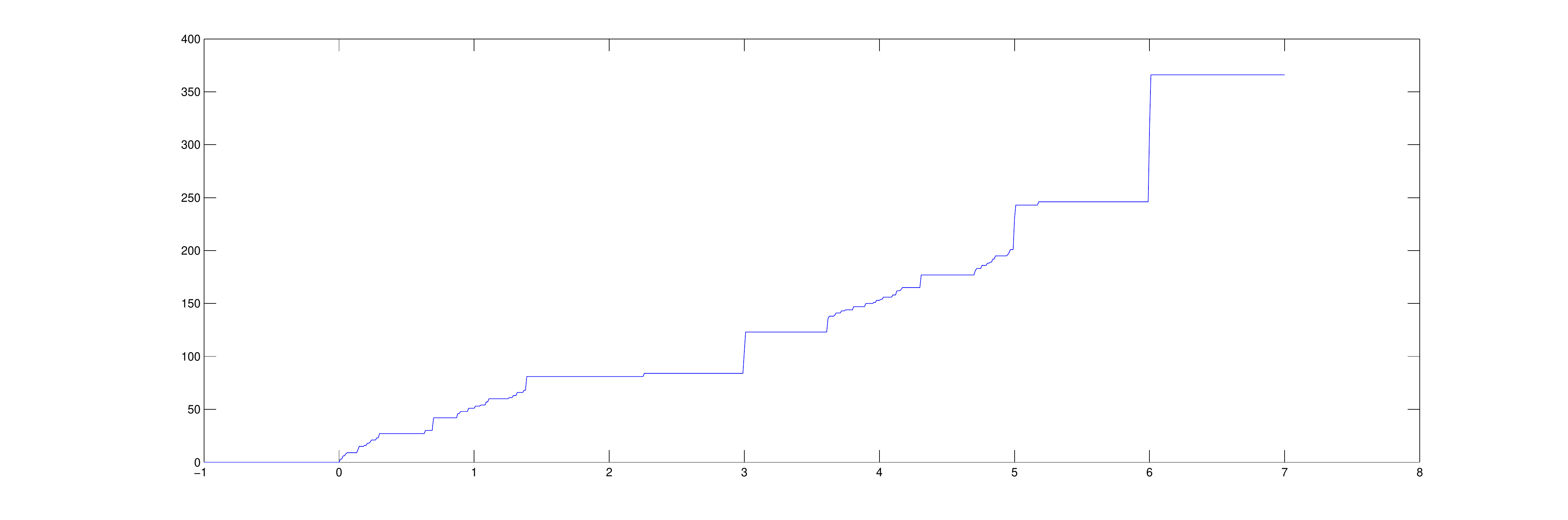}
\label{eigcountsg}}
\subfigure[Random Branching Tree $n=279, m=10$]{
\includegraphics[width=.45\linewidth]{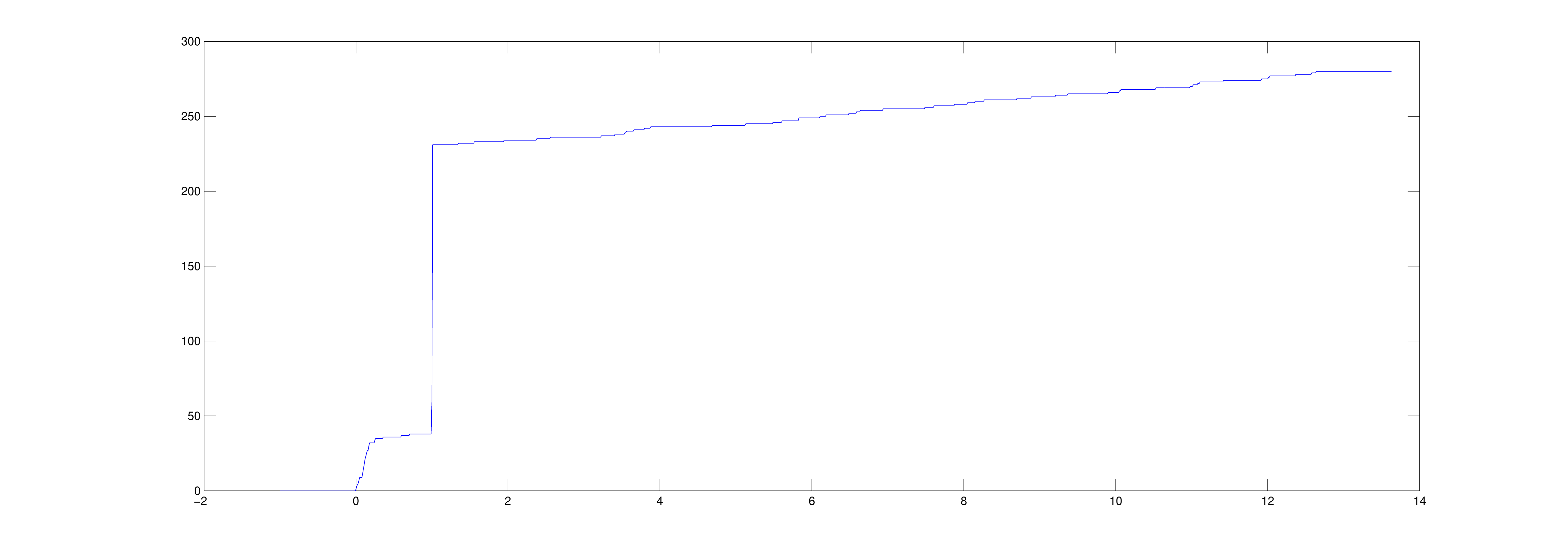}
\label{eigcounttree}}
\subfigure[Hexacarpet Level 3]{
\includegraphics[width=.45\linewidth]{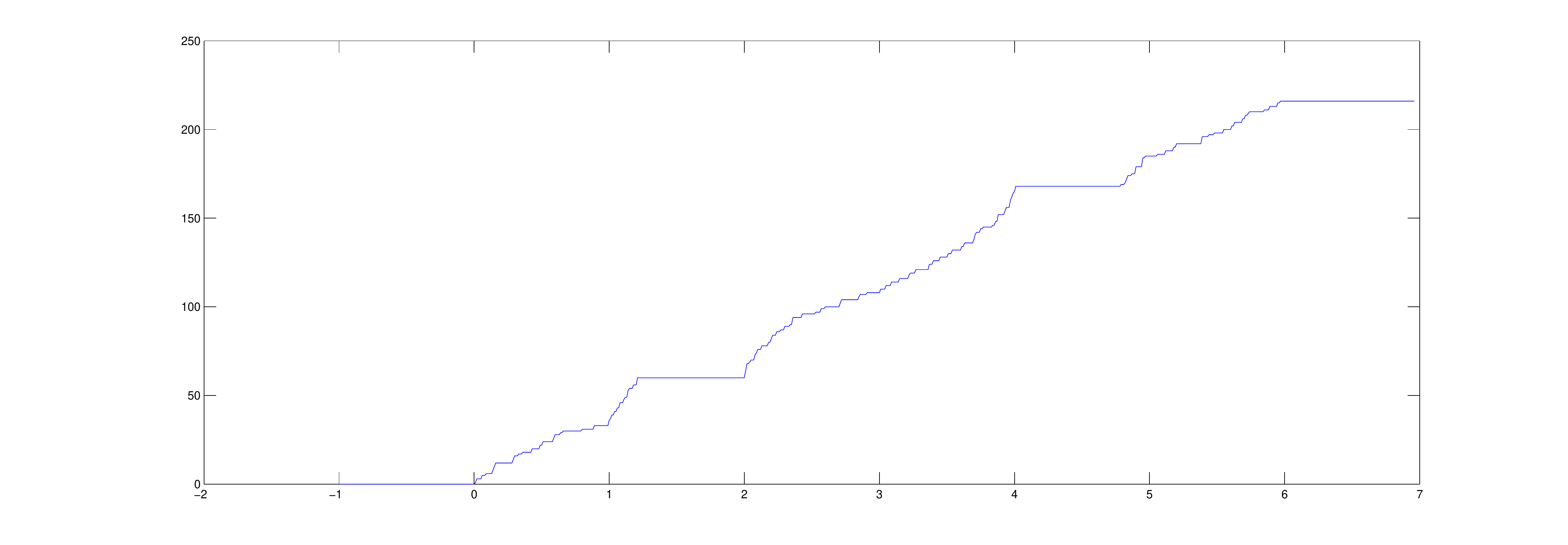}
\label{eigcounthexacarpet}}
\subfigure[Random Network $n=279, p=0.07$]{
\includegraphics[width=.45\linewidth]{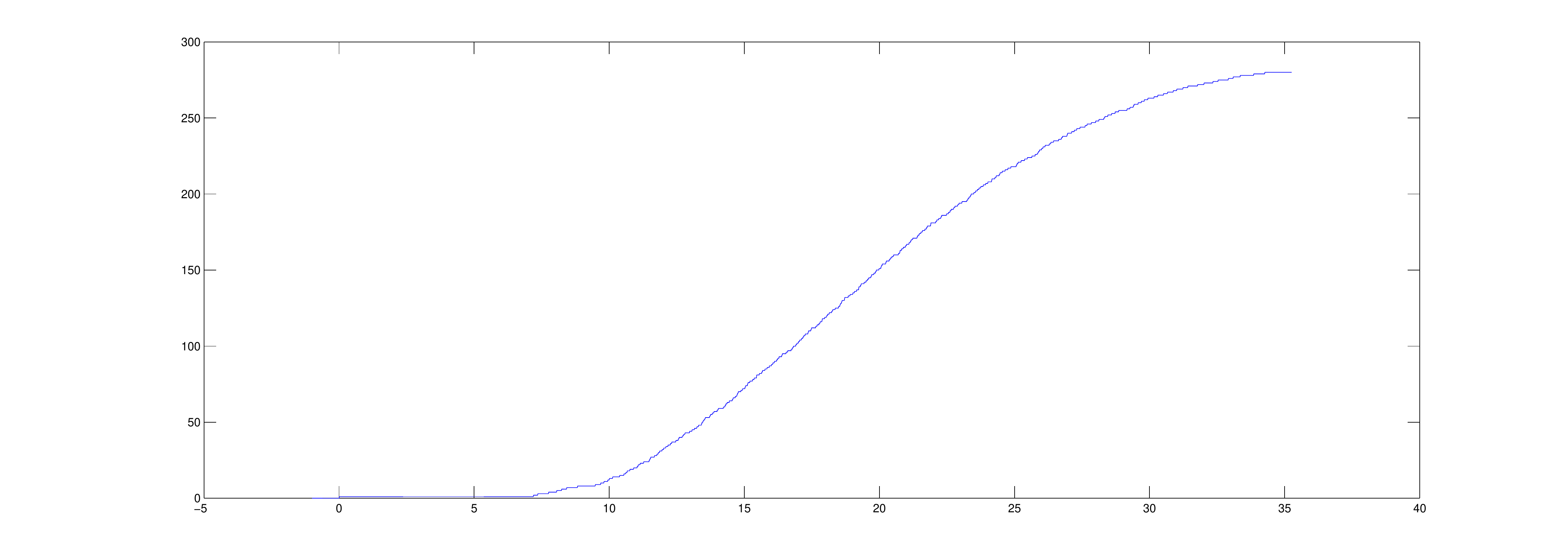}
\label{eigcountrandom}}
\subfigure[Octagasket, Level 3]{
\includegraphics[width=.45\linewidth]{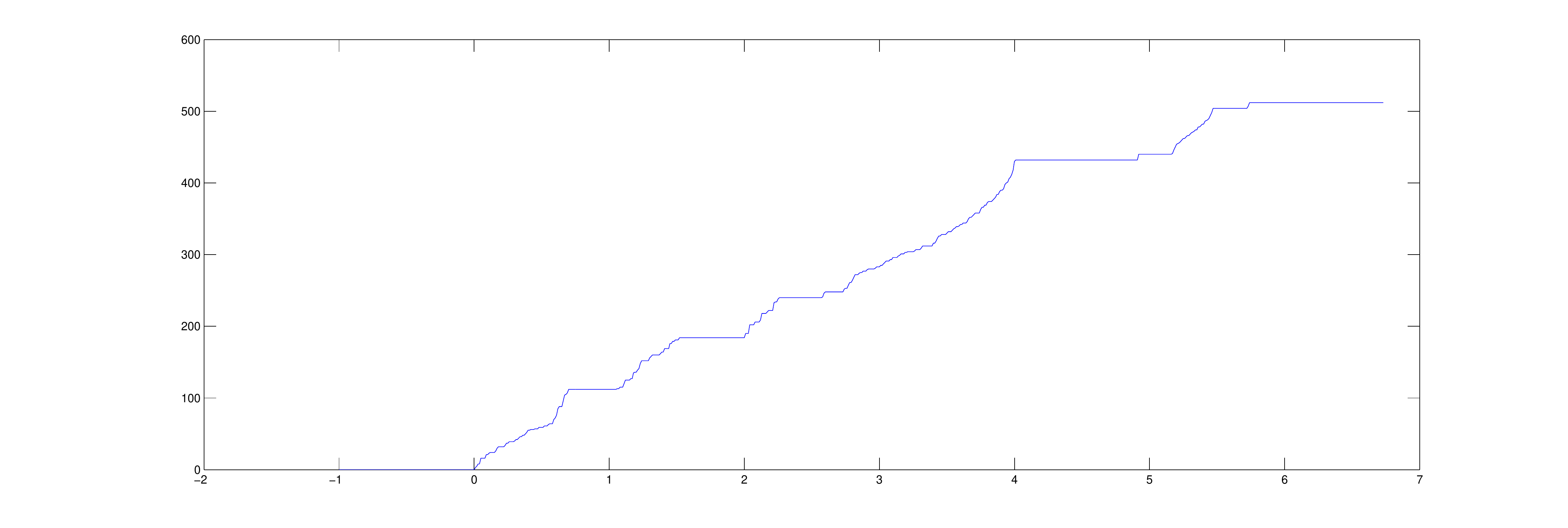}
\label{eigcountoctocarpet}}
\subfigure[Sierpinski Gasket Rewiring $p=0.15$]{
\includegraphics[width=.45\linewidth]{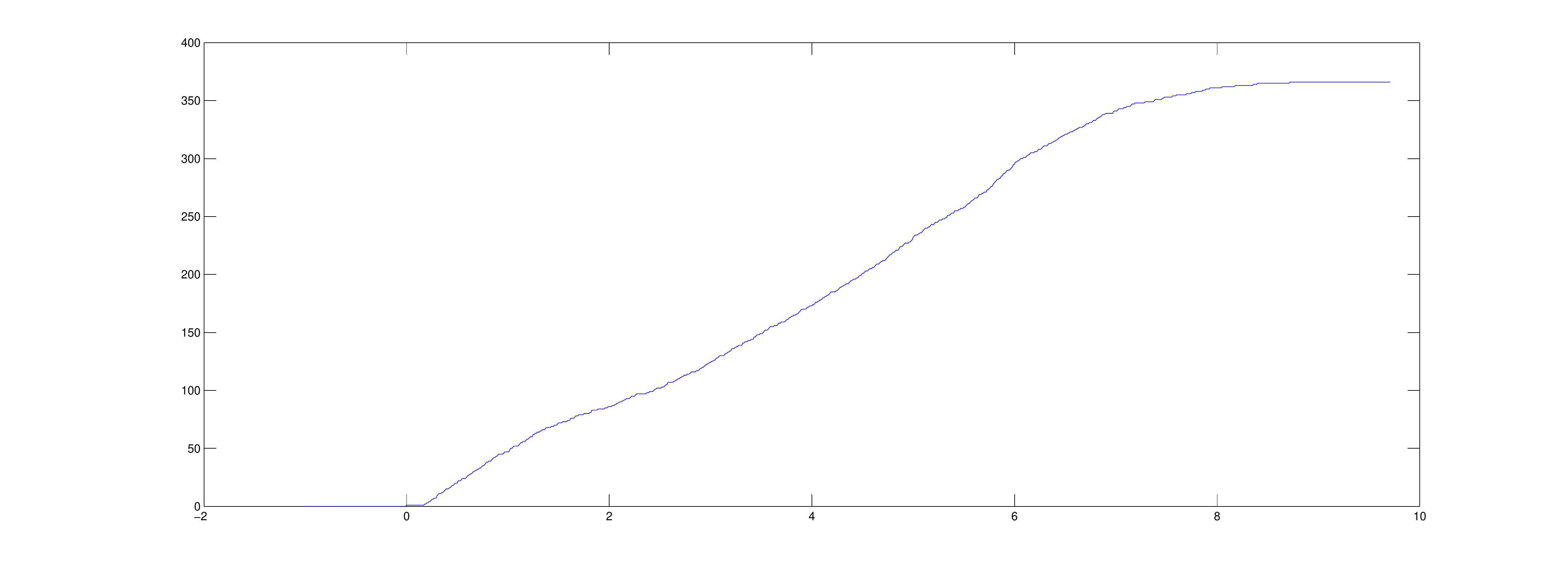}
\label{eigcountsgrewire}}
\subfigure[Sierpinski Carpet, Level 3]{
\includegraphics[width=.45\linewidth]{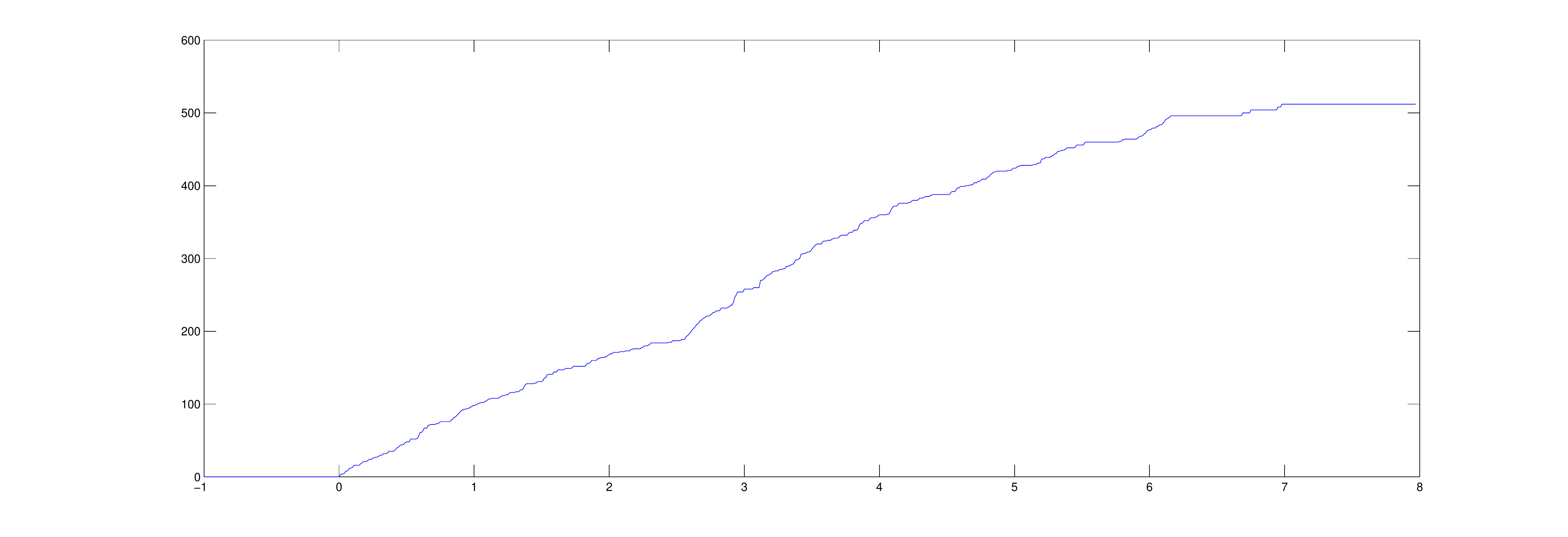}
\label{eigcountsierpinskicarpet}}
\label{eigcount}
\caption[Optional caption for list of figures]{The Eigenvalue Counting Function}
\end{figure}

	To begin our analysis, we applied the eigenvalue counting function to our Laplacian matrices.  The eigenvalue counting function is a cumulative distribution function on the spectrum of a matrix (see Methods \ref{methods3.1}).  It  computes the set of all eigenvalues of a given matrix (the spectrum), and counts the total number of eigenvalues less than the given input.  Plotting this function gives an expedient way to analyze how the graph of a given matrix should be generally organized, based on the spectrum of the graph Laplacian \cite{DKC}.  Figure \ref{eigcount} shows the plots of the eigenvalue counting function on a variety of our Laplacian matrices.

\noindent

	Upon examination of these graphs, a few patterns immediately emerge.  The first is the presence of step-like portions of those graphs corresponding to known fractals (Fig. \ref{eigcountsg}, \ref{eigcounthexacarpet}, \ref{eigcountoctocarpet}, \ref{eigcountsierpinskicarpet}).  These sections of slope-zero correspond to gaps in the spectrum of the graph.  This feature is closely linked to self-similar fractal structures \cite{ZD}.  Although the eigenvalue counting function plot of the \emph{C. elegans} brain (Fig. \ref{eigcountworm}) does not show these definitive spectral gaps, this does not conclusively eliminate the possibility of self-similar structures existing within the neural network.  At the same time, however, this does indicate that the nematode brain is \emph{not} strictly fractal-like in structure.
	
	Another interesting pattern appears in the graph corresponding to the random-branching tree (Fig \ref{eigcounttree}).  There is a large vertical jump at $x=1$, with a change on the y-axis of approximately 200.  This indicates that for this graph Laplacian, the eigenvalue 1 occurs with extremely high multiplicity.  This is caused by the nature of the tree's structure.  With a finite number of points (in this case 279) a large number of these points are endpoints: vertices at which no further branching occurs.  These points are only connected to one other: their ``parent" vertex.  The lack of this spectral ``jump" in the \emph{C. elegans} brain indicates that there are very few singly-connected endpoints in the system.  Rather, the neural network is highly inter-connected, with structural properties baring little resemblance to the branches of a tree. 
		
		On the other hand, the eigenvalue counting patterns of the \emph{C. elegans} neural network do resemble those of the random network (Fig. \ref{eigcountrandom}) and the rewired Sierpinski Gasket (Fig \ref{eigcountsgrewire}).  While a similarity in these patterns  (as opposed to the aforementioned dissimilarities) cannot conclusively point to similar structural properties, these eigenvalue counting functions are a prerequisite to the next step in our analysis, Weyl ratios. 
		
\subsection{Weyl Ratios}

\begin{figure}[ht]
\centering
\subfigure[\emph{C. elegans} neural network]{
\includegraphics[width=.45\linewidth]{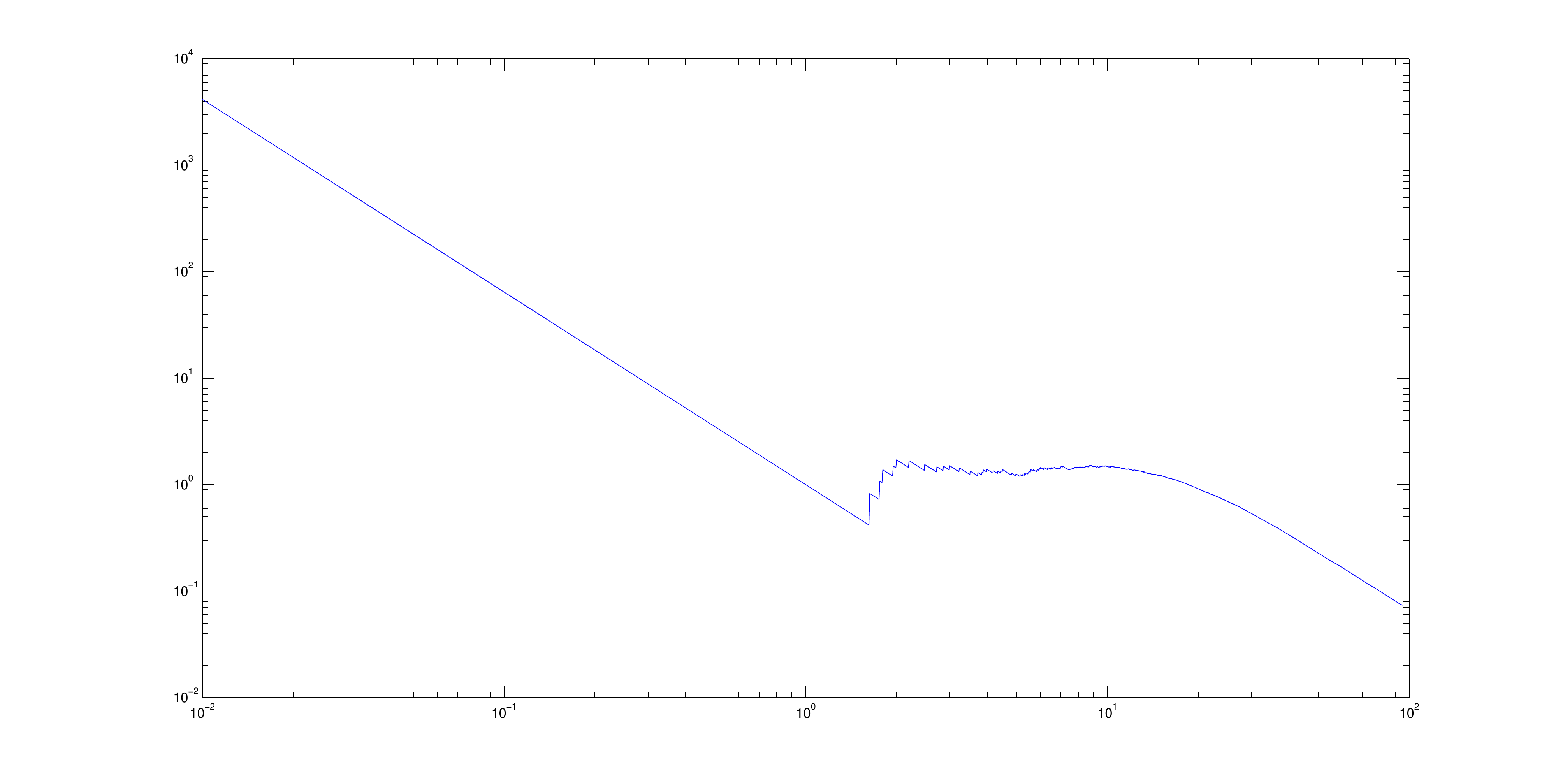}
\label{Weylratioworm}}
\subfigure[Sierpinski Gasket, Level 5]{
\includegraphics[width=.45\linewidth]{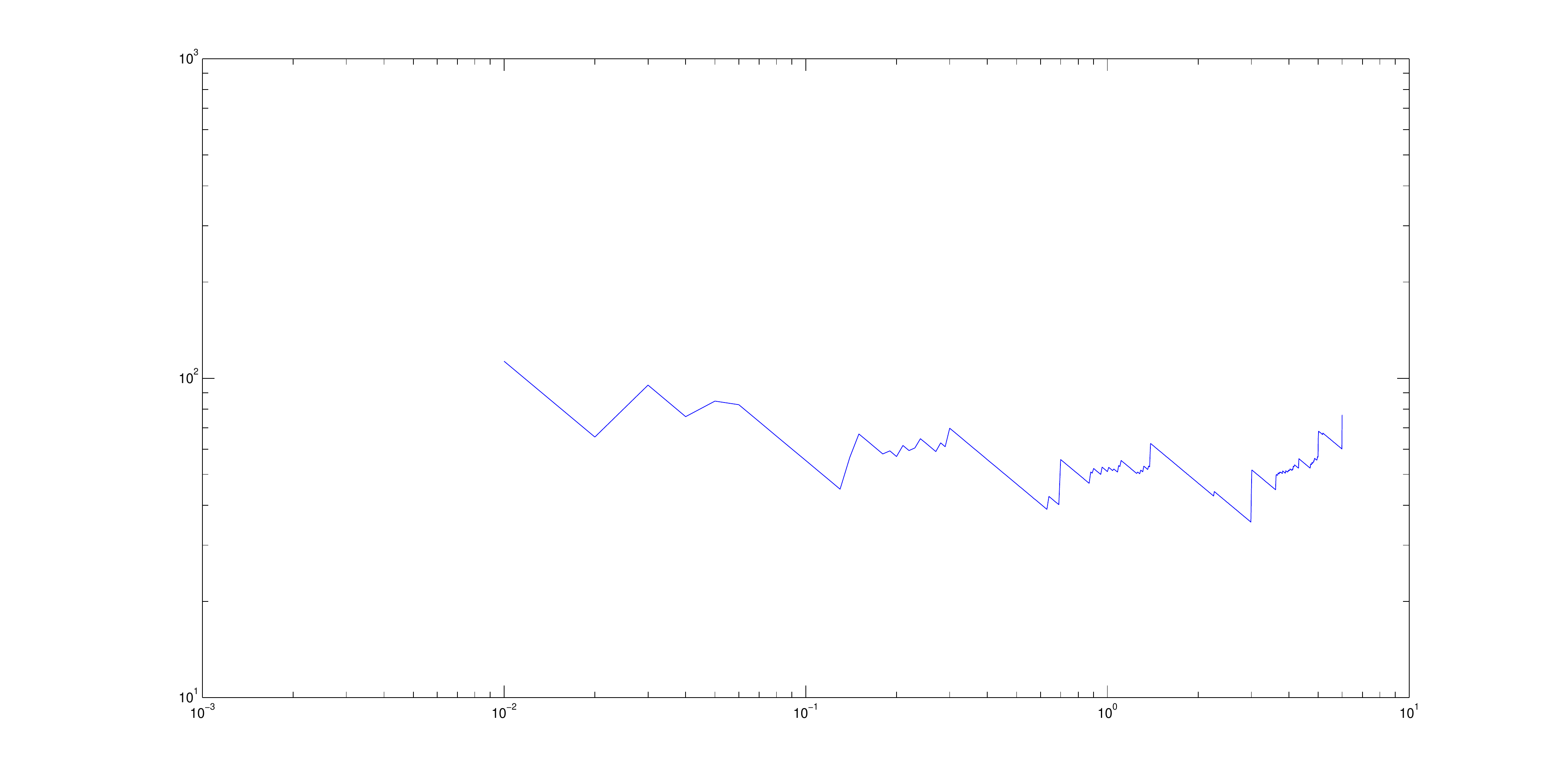}
\label{Weylratiosg}}
\subfigure[Random-Branching Tree $n=279, m=10$]{
\includegraphics[width=.45\linewidth]{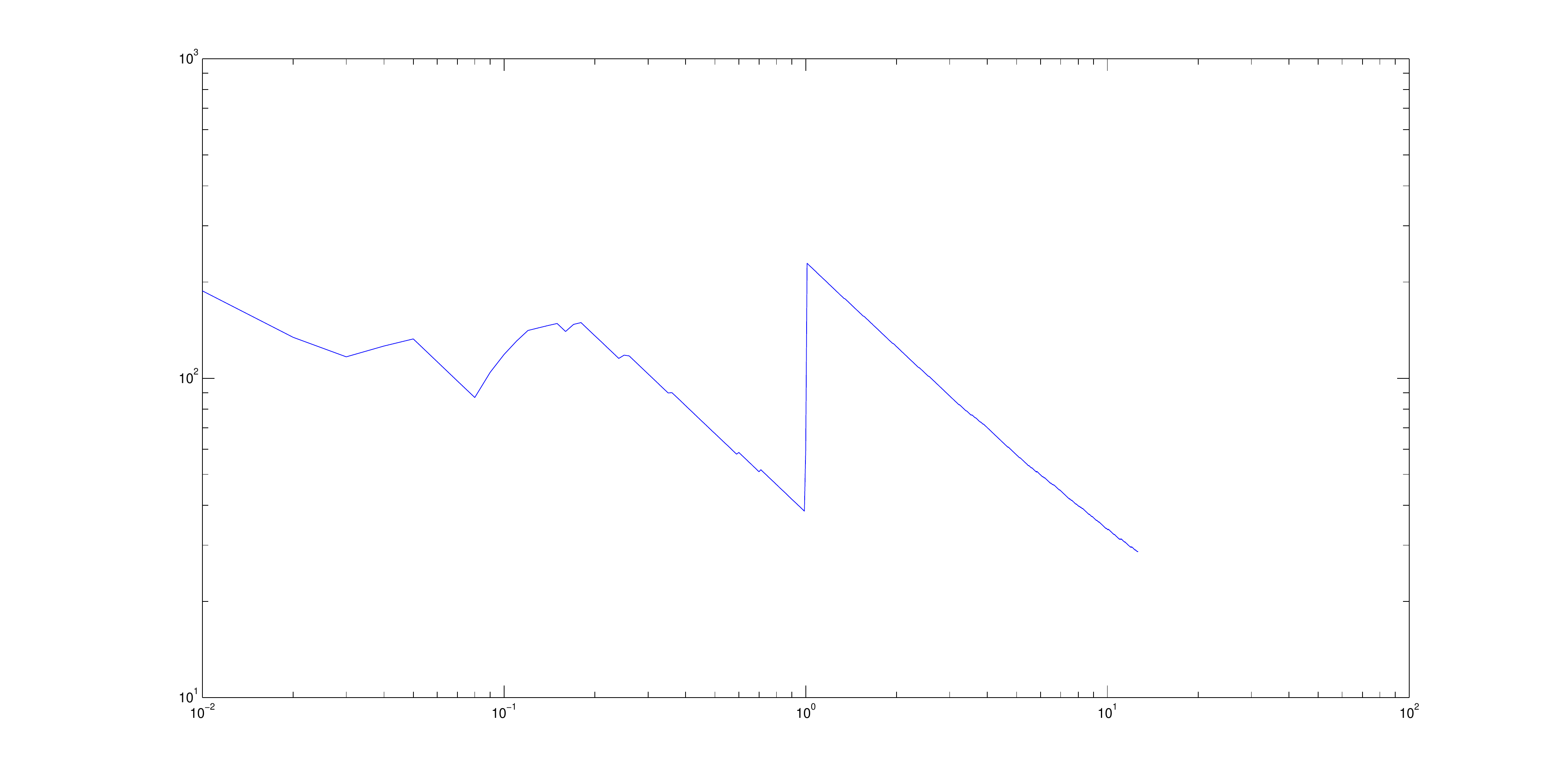}
\label{Weylratiotree}}
\subfigure[Hexacarpet Level 3]{
\includegraphics[width=.45\linewidth]{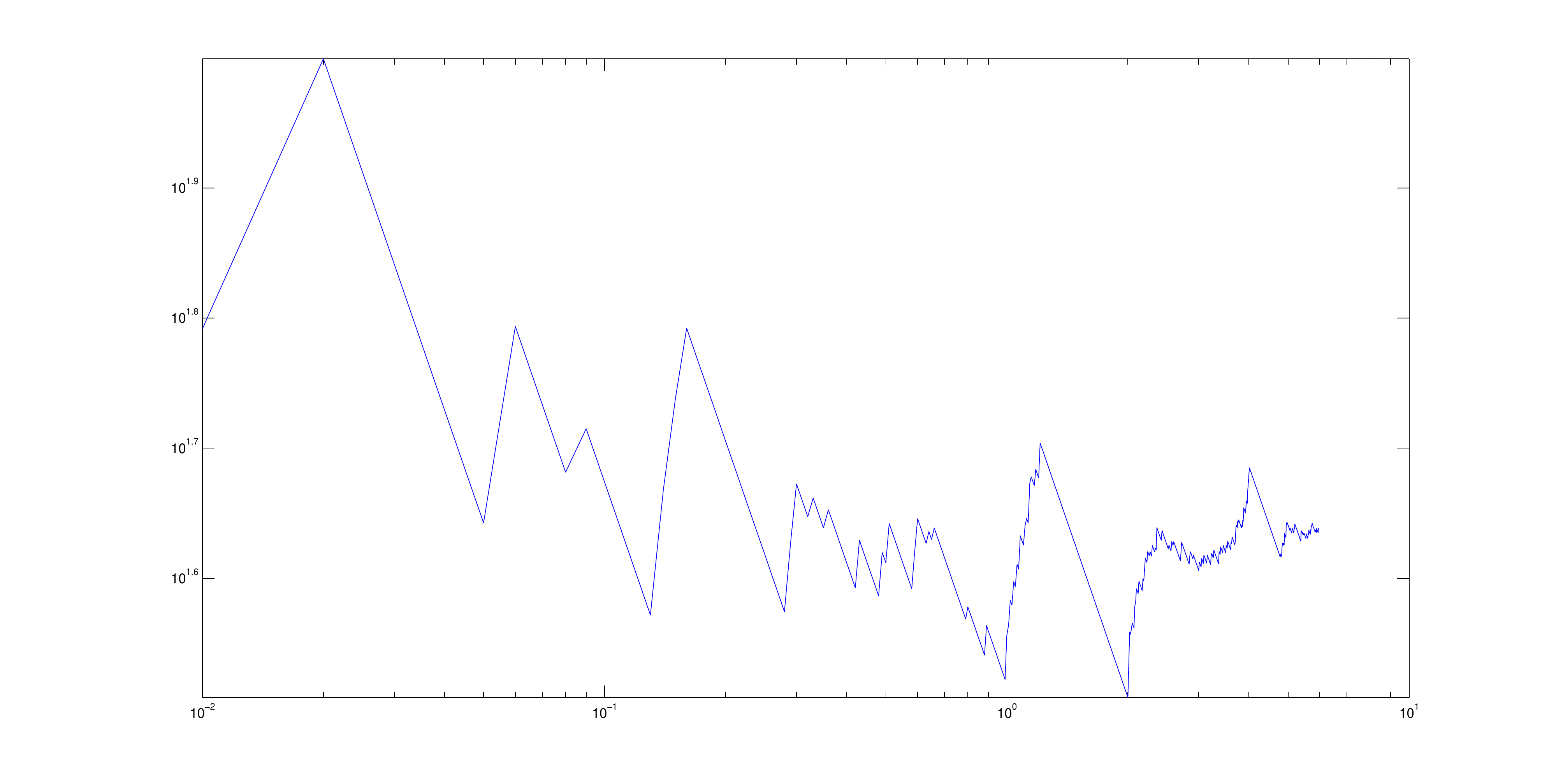}
\label{Weylratiohexacarpet}}
\subfigure[Random Network $n=279, p=0.07$]{
\includegraphics[width=.45\linewidth]{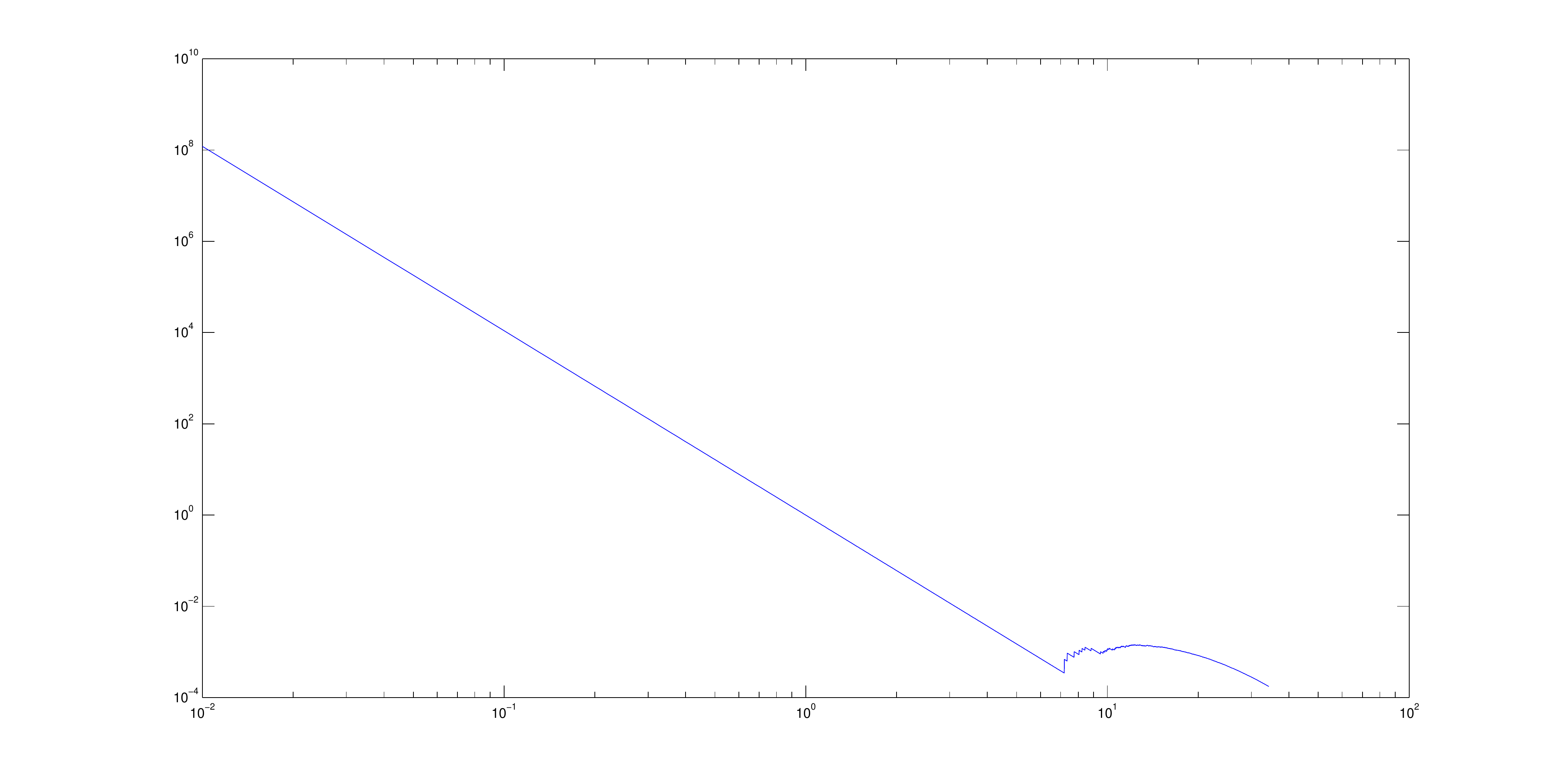}
\label{Weylratiorandom}}
\subfigure[Octagasket, Level 3]{
\includegraphics[width=.45\linewidth]{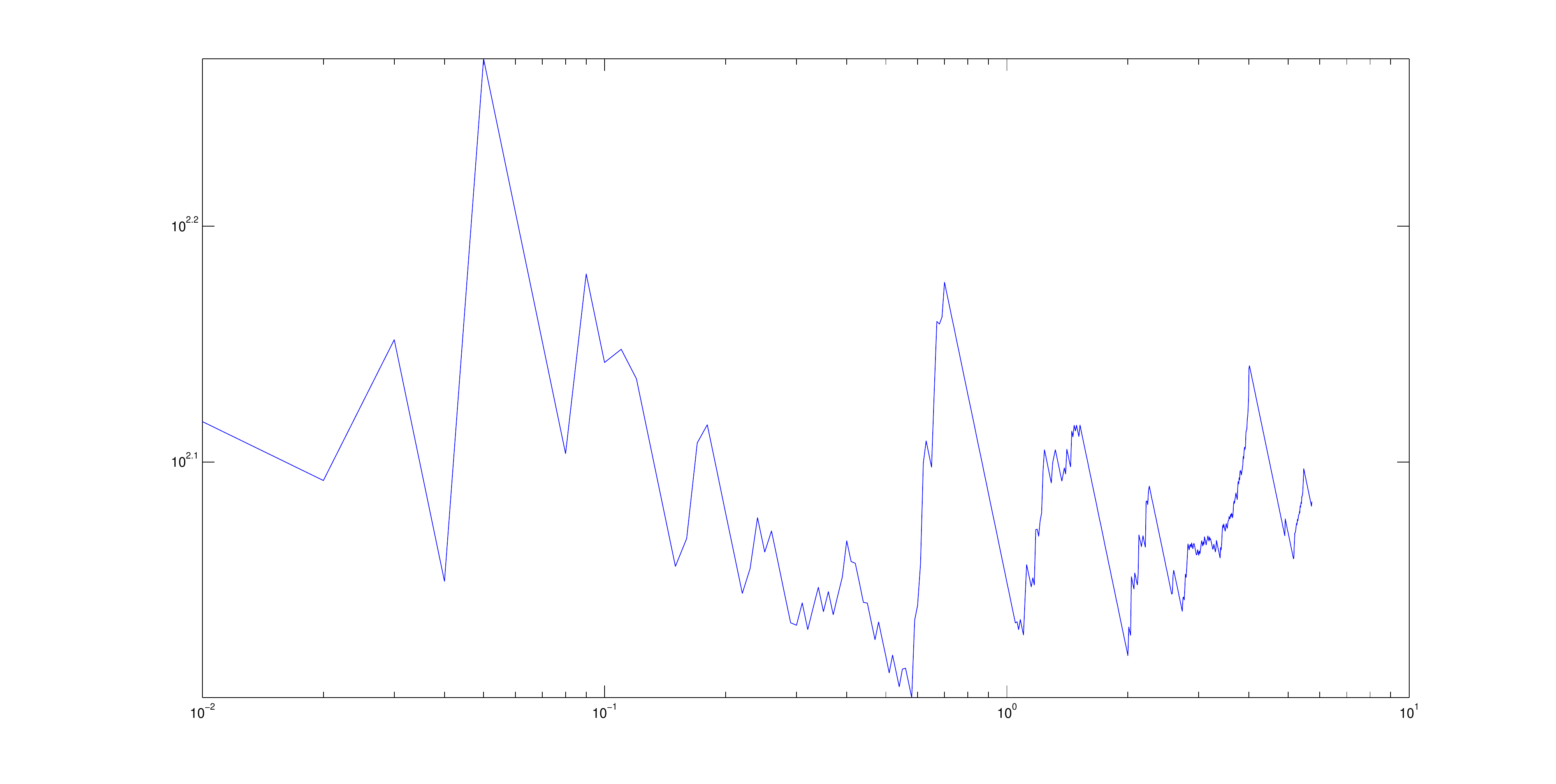}
\label{Weylratiooctocarpet}}
\subfigure[Sierpinski Gasket Rewiring $p=0.15$]{
\includegraphics[width=.45\linewidth]{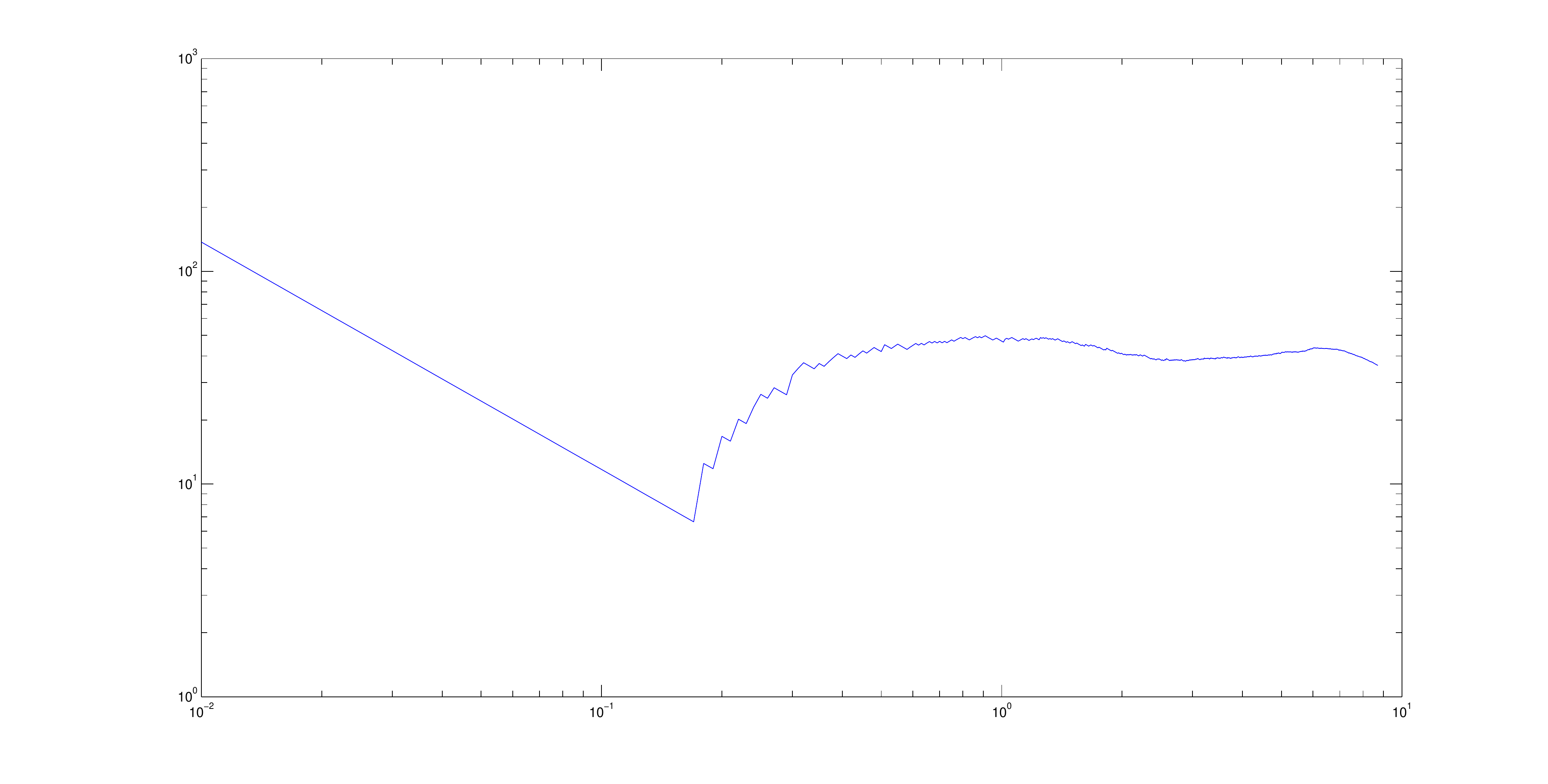}
\label{Weylratiosgrewire}}
\subfigure[Sierpinski Carpet, Level 3]{
\includegraphics[width=.45\linewidth]{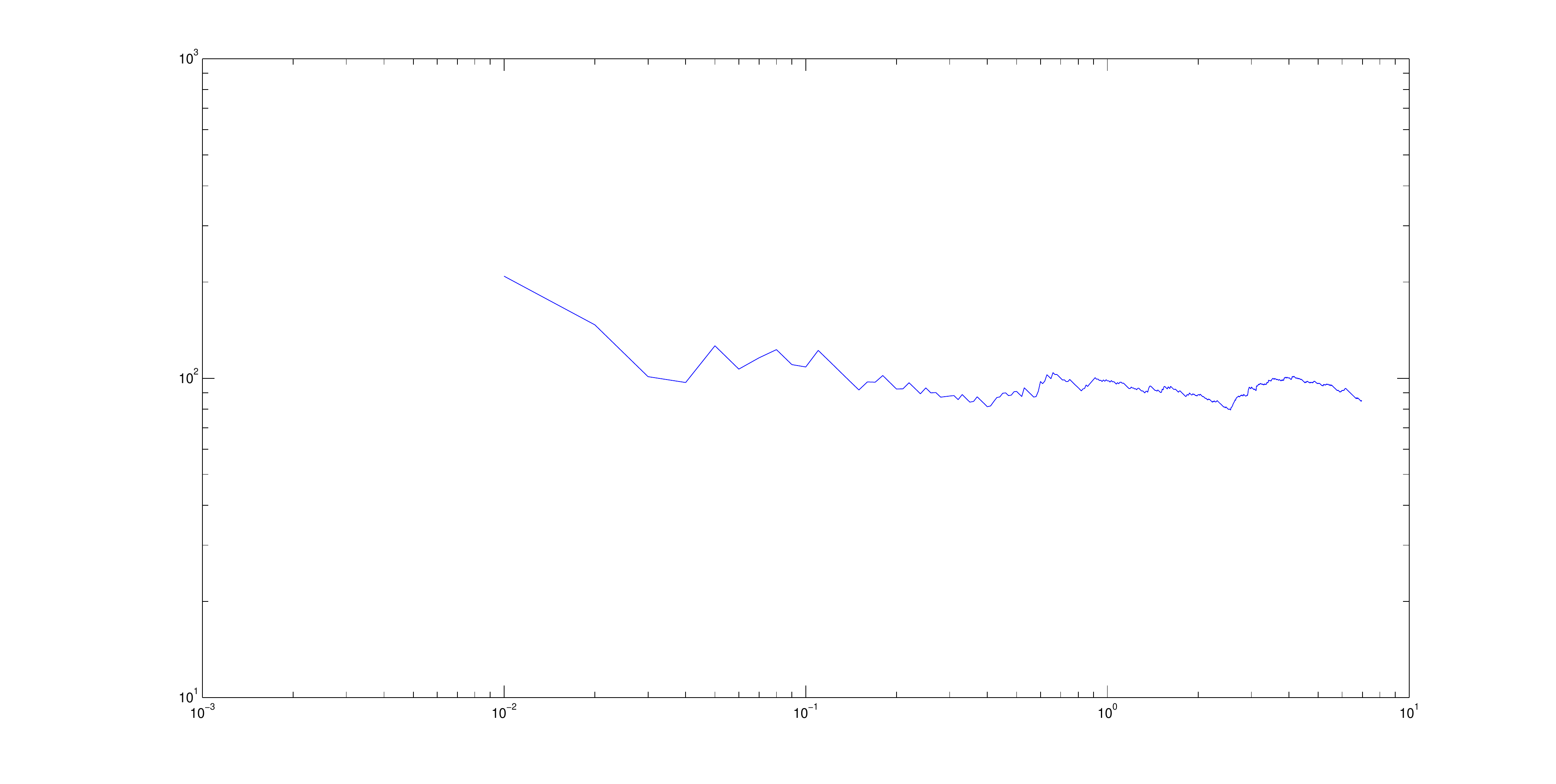}
\label{Weylratiosierpinskicarpet}}
\label{Weylratio}
\caption[Optional caption for list of figures]{Weyl Ratios}
\end{figure}

	In order to find Weyl ratios for our network, we plotted the respective eigenvalue counting functions on a log-log scale.  We then searched for an interval on which this graph is roughly linear.  We determined a line of best fit for this section, and found its slope, $\alpha$. We then used this $\alpha$ to produce a Weyl ratio for each graph (see Methods \ref{methods3.1}).  A Weyl ratio is essentially a rescaling of points in our eigenvalue counting graph, providing a more revealing visualization.  Figure \ref{Weylratio} shows the Weyl ratios for each of our Laplacian matrices.  Periodicity in the Weyl ratio plot suggests the presence of self-similar structures.  For more on Weyl ratio analysis of known fractals, see \cite{BHS26}. 
	 
\noindent

	As expected, the Weyl ratios of known self-similar fractals (Fig. \ref{Weylratiosg}, \ref{Weylratiohexacarpet}, \ref{Weylratiooctocarpet}, \ref{Weylratiosierpinskicarpet}) show a high degree of organization.  That of the Sierpinski gasket in particular (Fig. \ref{Weylratiosg}), shows unmistakable periodicity.  Although not as readily noticeable, clear patterns exist in the Weyl ratios of other fractals as well.
	 
	It should be noted that the Weyl ratio graph of the branching-tree (Fig. \ref{Weylratiotree}) is different from those of other networks.  Although the tree structure lacks the high ordering of known fractals, there is a clear distinction between this special case and all other networks: what we will call ``looping".  In fractals, random networks, and the \emph{C. elegans} brain alike, the vertices are highly interconnected.  Many cyclic paths exist which allow a signal to arrive back at a starting vertex by traveling through a series of other vertices.  Trees, on the other hand, lack this feature: only one path exists between any two points.  Not only does this create a unique Weyl ratio pattern, but it also suggests that neural networks are not strictly branching structures, as one might expect.
	 
	While the Weyl ratio pattern of the \emph{C. elegans} brain ( Fig. \ref{Weylratioworm}) is dissimilar from that of a branching tree, it also differs greatly from those of known fractals.  While several cases of slight periodicity could be argued for, this evidence is not definitive enough to indicate concrete self-similarity in the \emph{C. elegans} neural network.  However, there does exist similarity between the Weyl ratio patterns of the C. elegans neural network, the randomly generated network (Fig. \ref{Weylratiorandom}), and the rewiring of the Sierpinski Gasket (Figure \ref{Weylratiosgrewire}).  (It should be noted that many random networks and SG rewirings were generated, these two examples were chosen as representatives). Although the significance of examining a "rewired" fractal structure will be explained in Section \ref{smallworld} of this paper, it is important to note that this likeness in Weyl ratio patterns can suggest a structural similarity.  While it seems counterintuitive that the nematode brain would be a random arrangement of neurons, the evidence found in the Weyl ratios alone cannot conclude that this is (or is not) indeed the case.  Our further analysis will make distinctions between these 3 networks.
 
\subsection{The Eigen-Projection Method}

	Next, we wanted a way to visualize the neural network.  We did this by embedding a graph of all neurons in Euclidean space via the eigen-projection method explained in \cite{K27}, similar to those processes described in \cite{M28, PST29}.  This spectral approach to visualizing graphs utilizes the eigenfunctions of degree-normalized Laplacian matrices (see Methods \ref{methods3.2}).

	The eigen-projection method plots the vertices of a graph using the eigenfunctions of its Laplacian matrix as a coordinate system (it is also referred to as `plotting in eigenvector coordinates').  This method essentially projects the graphs into a smaller Euclidean space, typically either $\mathbb R^2$ or $\mathbb R^3$, using appropriate eigenfunctions.  See Methods \ref{methods3.3} for a more rigorous description.
	
	After embedding each vertex in either 2- or 3-dimensional space, we then represented neuronal connections (or network connections) with line segments between the appropriate points.  In the case of our \emph{C. elegans} brain diagram, we also used the same color-coding as \cite{VPC25}: where red represents sensory neurons, green are motor neurons, and blue indicates interneurons.  Lastly, we labeled the points with the corresponding neuron name abbreviations.  This was done using a slight variation of the VISUALIZE program used by Chklovskii and Varshney, available at \cite{PVC30}.
	
\noindent
\begin{figure}[!]
\centering
\subfigure[\emph{C. elegans} neural network, $(\varphi_2,\varphi_3)$]{
\includegraphics[width=.45\linewidth]{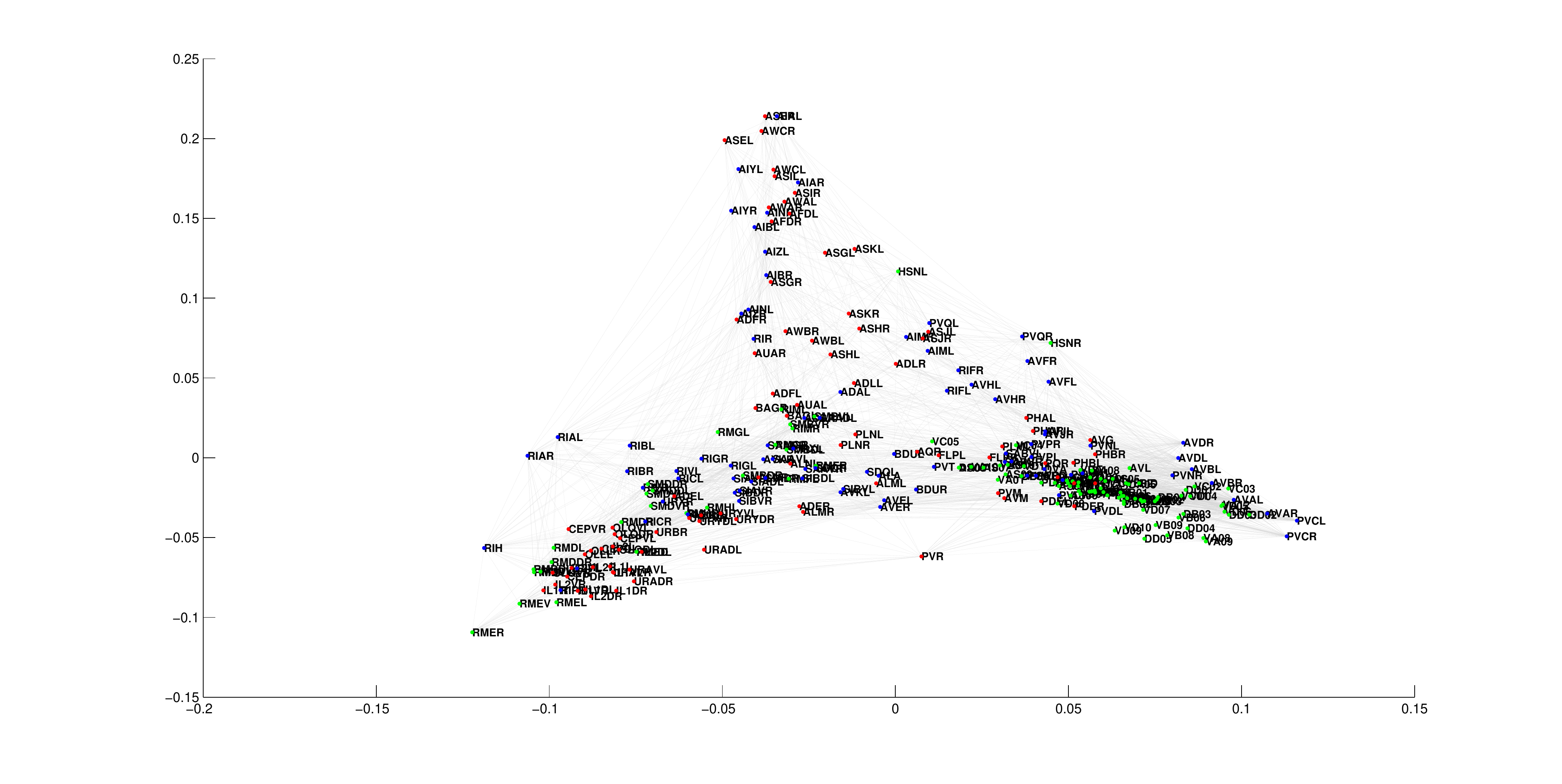}
\label{eigprojectionworm}}
\subfigure[\emph{C. elegans} neural network, $(\varphi_2,\varphi_3,\varphi_4)$]{
\includegraphics[width=.45\linewidth]{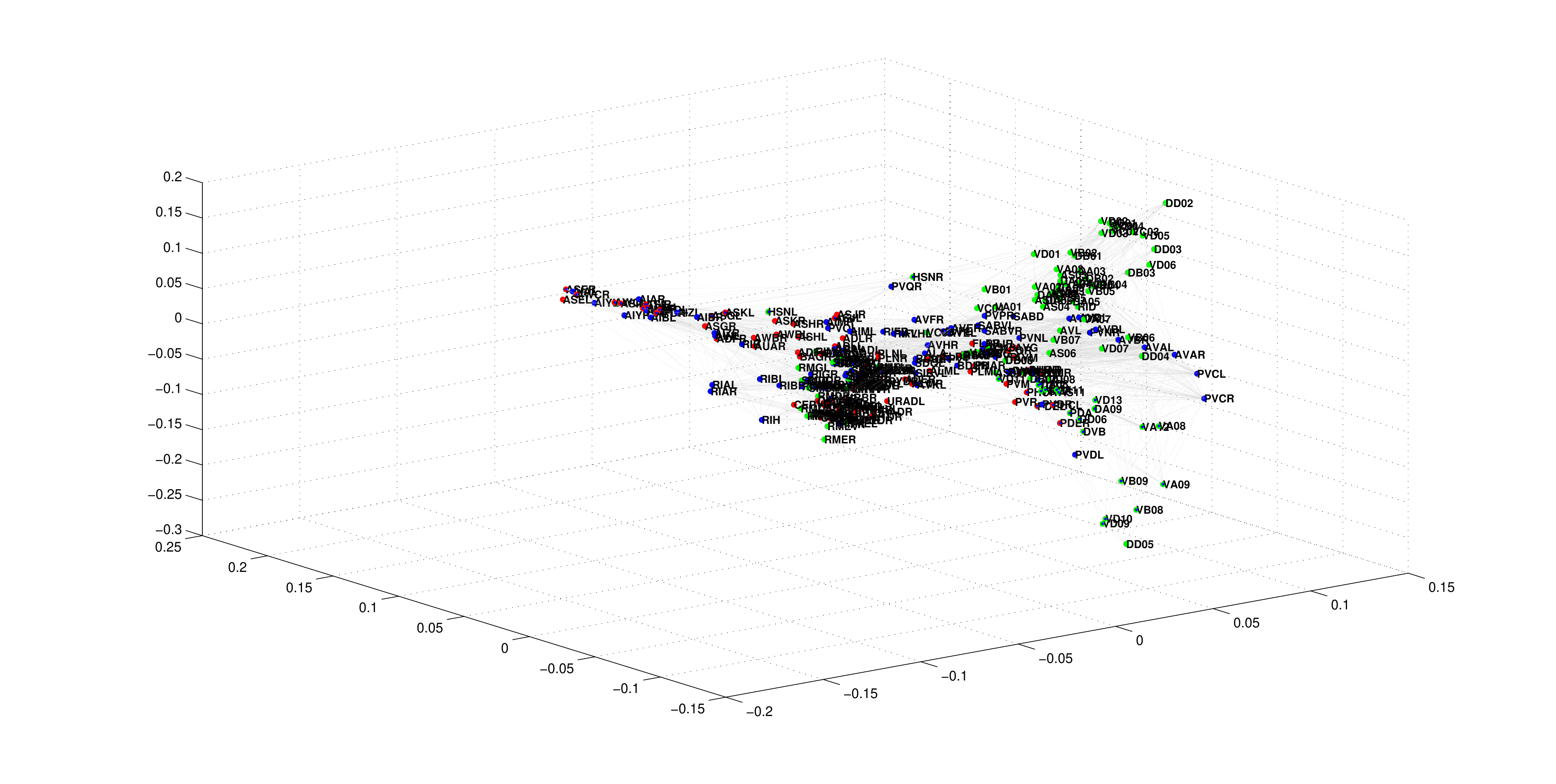}
\label{eigprojectionworm3D}}
\subfigure[Sierpinski Gasket, Level 5, $(\varphi_2,\varphi_3)$]{
\includegraphics[width=.45\linewidth]{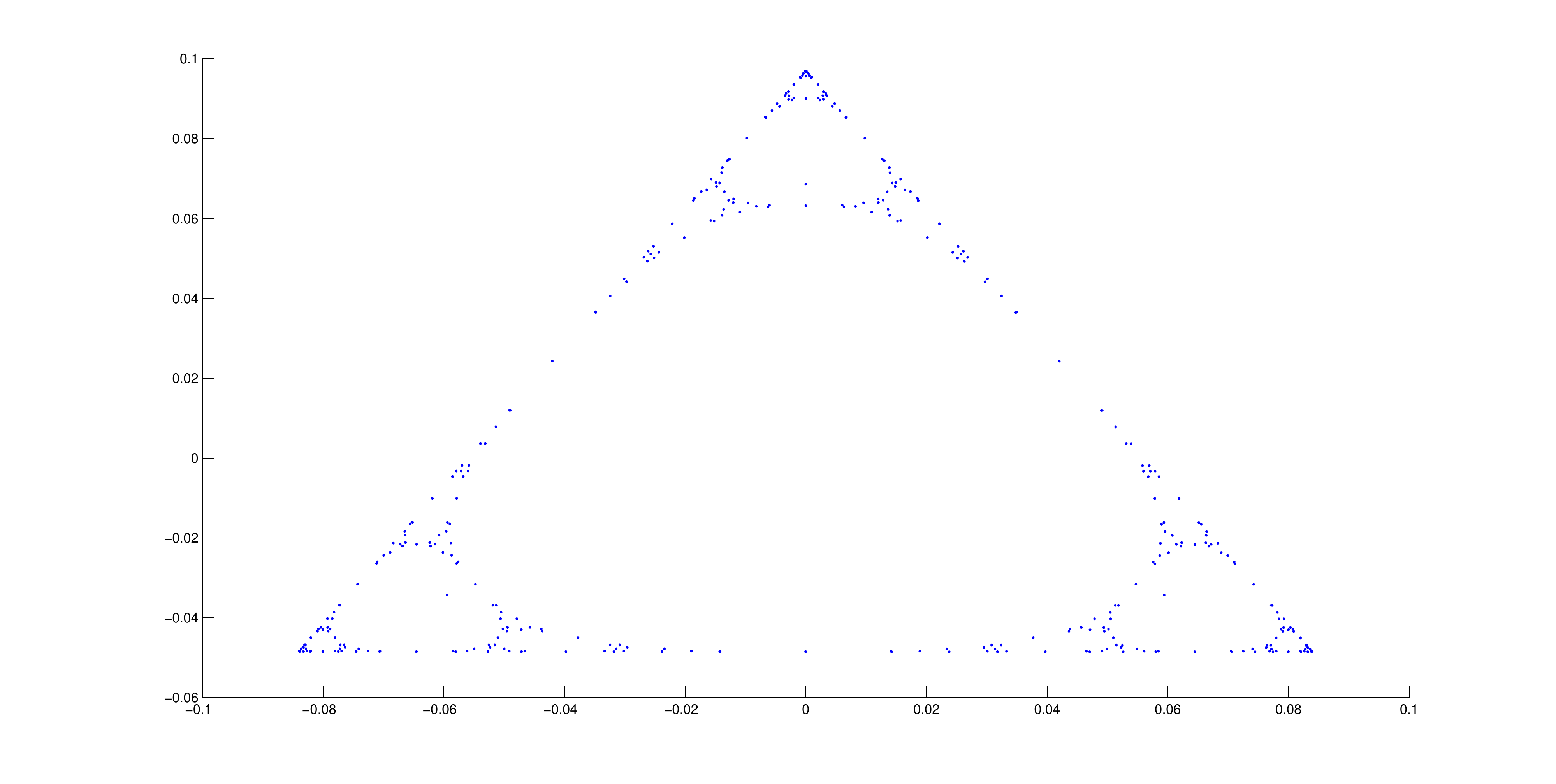}
\label{eigprojectionsg}}
\subfigure[Sierpinski Gasket, Level 5, $(\varphi_2,\varphi_3,\varphi_4)$]{
\includegraphics[width=.45\linewidth]{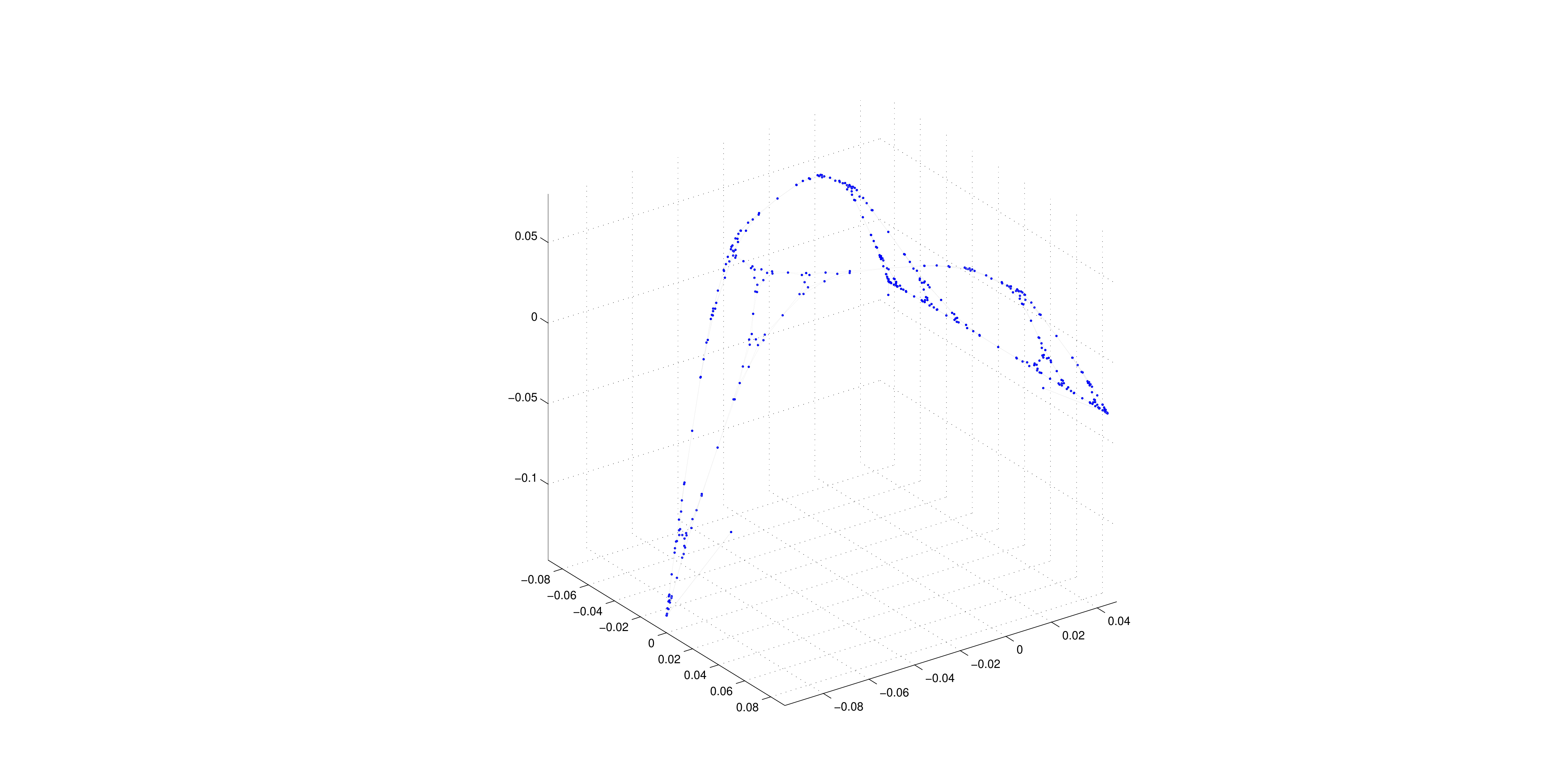}
\label{eigprojectionsg3D}}
\subfigure[Random Network $n=279, p=0.07$, $(\varphi_2,\varphi_3)$]{
\includegraphics[width=.45\linewidth]{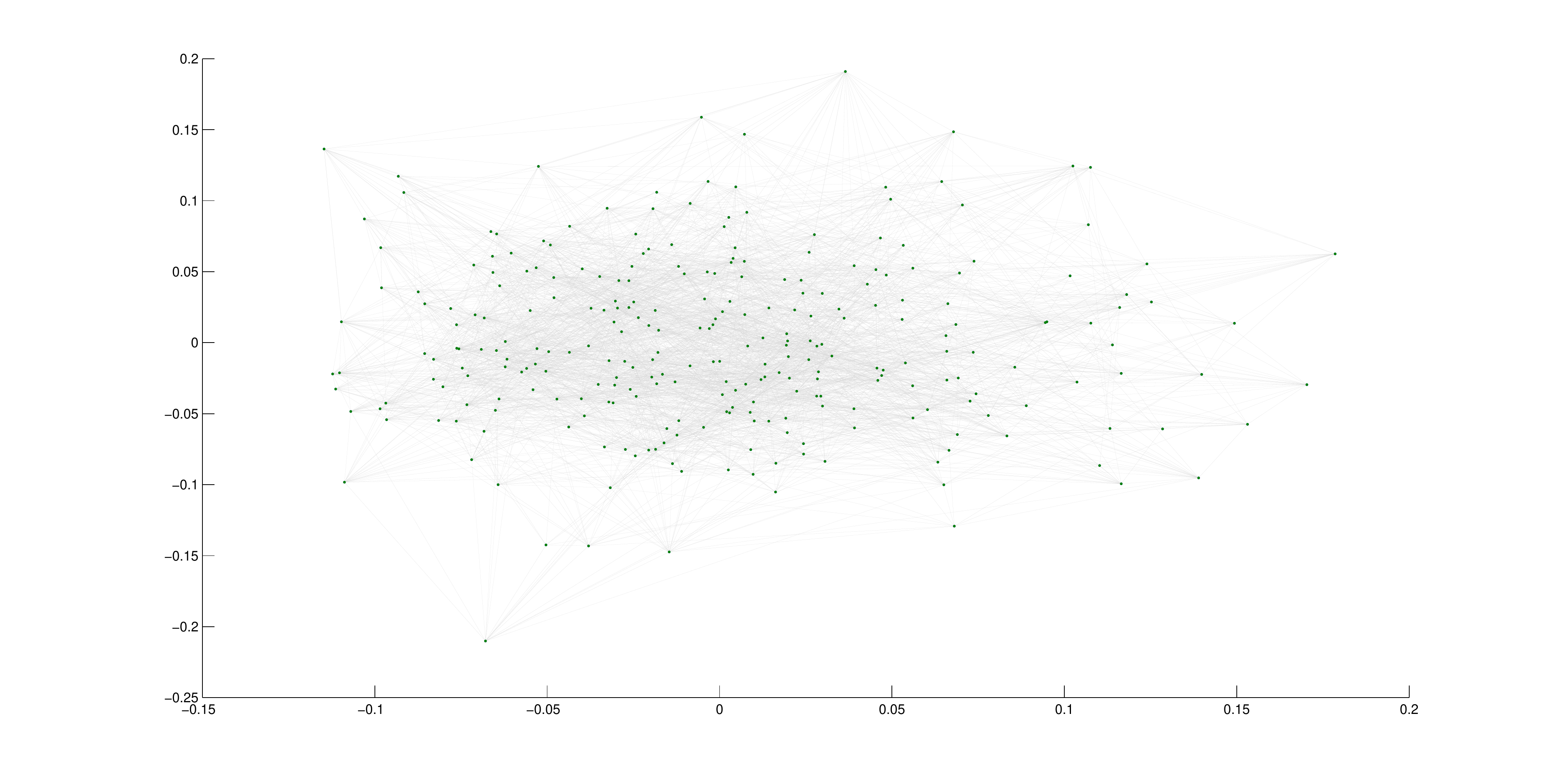}
\label{eigprojectionrandom}}
\subfigure[Random Network $n=279, p=0.07$, $(\varphi_2,\varphi_3,\varphi_4)$]{
\includegraphics[width=.45\linewidth]{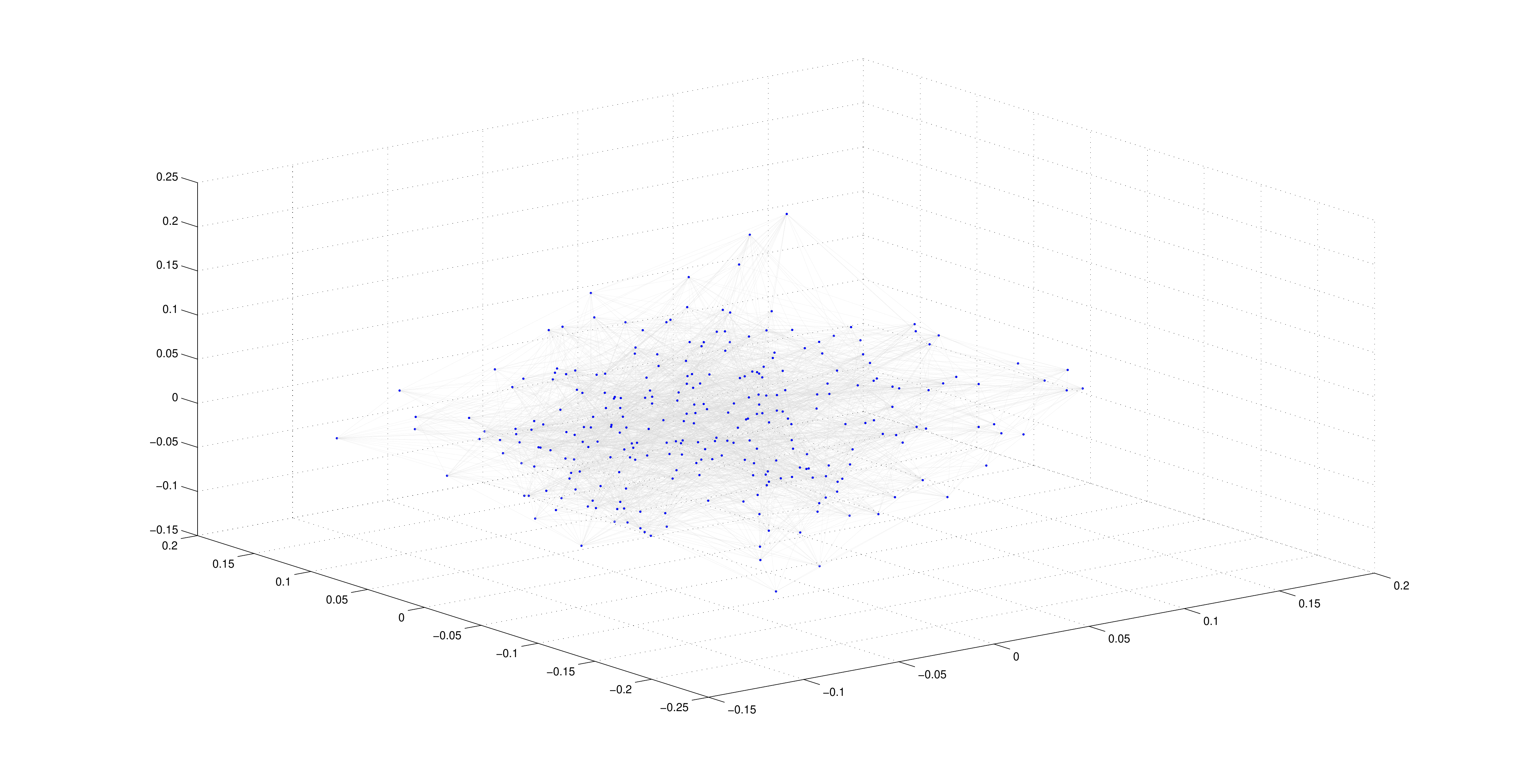}
\label{eigprojectionrandom3D}}
\subfigure[Sierpinski Gasket Rewiring $p=0.15$, $(\varphi_2,\varphi_3)$]{
\includegraphics[width=.45\linewidth]{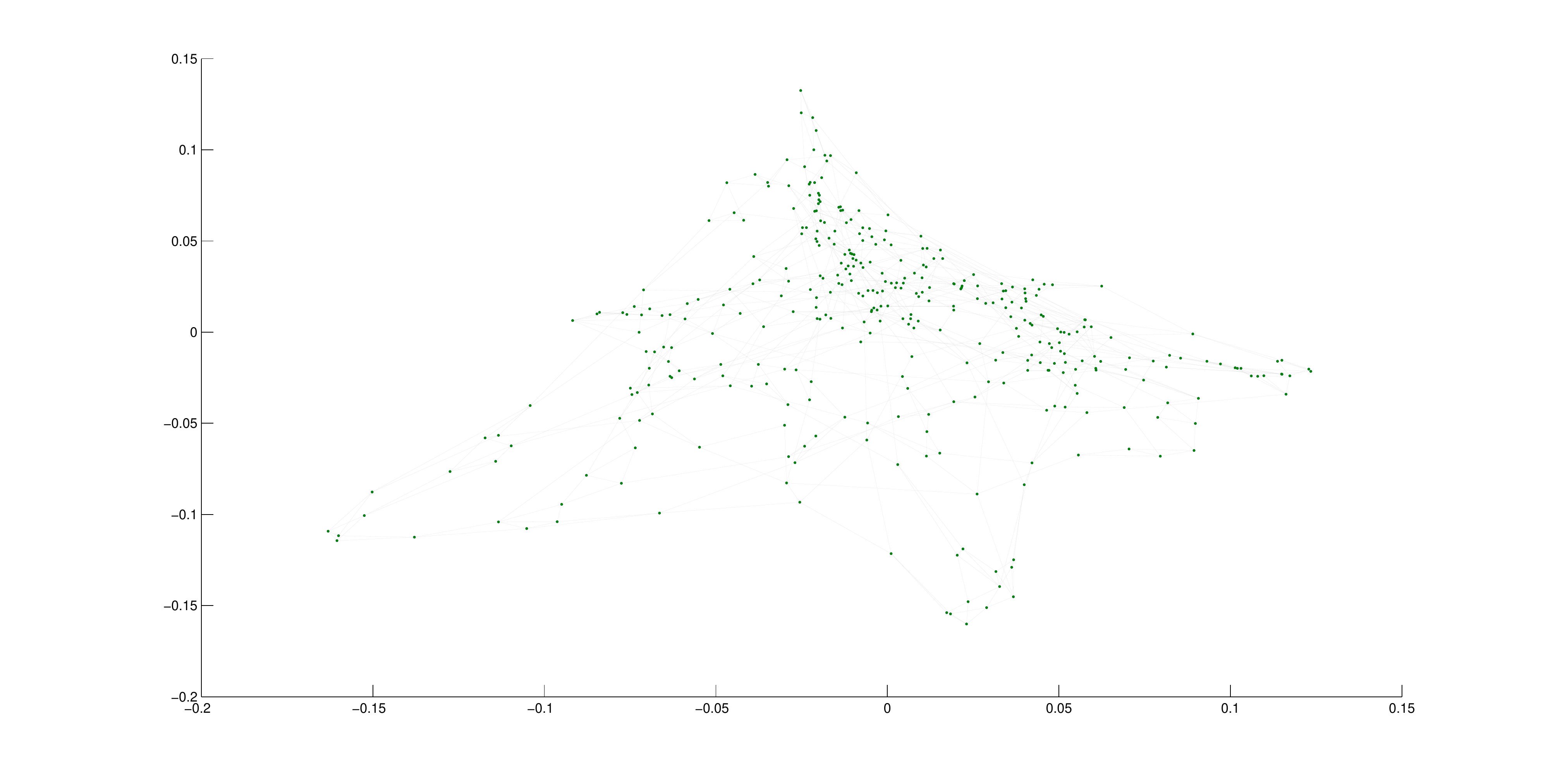}
\label{eigprojectionsgrewire}}
\subfigure[Sierpinski Gasket Rewiring $p=0.15$, $(\varphi_2,\varphi_3,\varphi_4)$]{
\includegraphics[width=.45\linewidth]{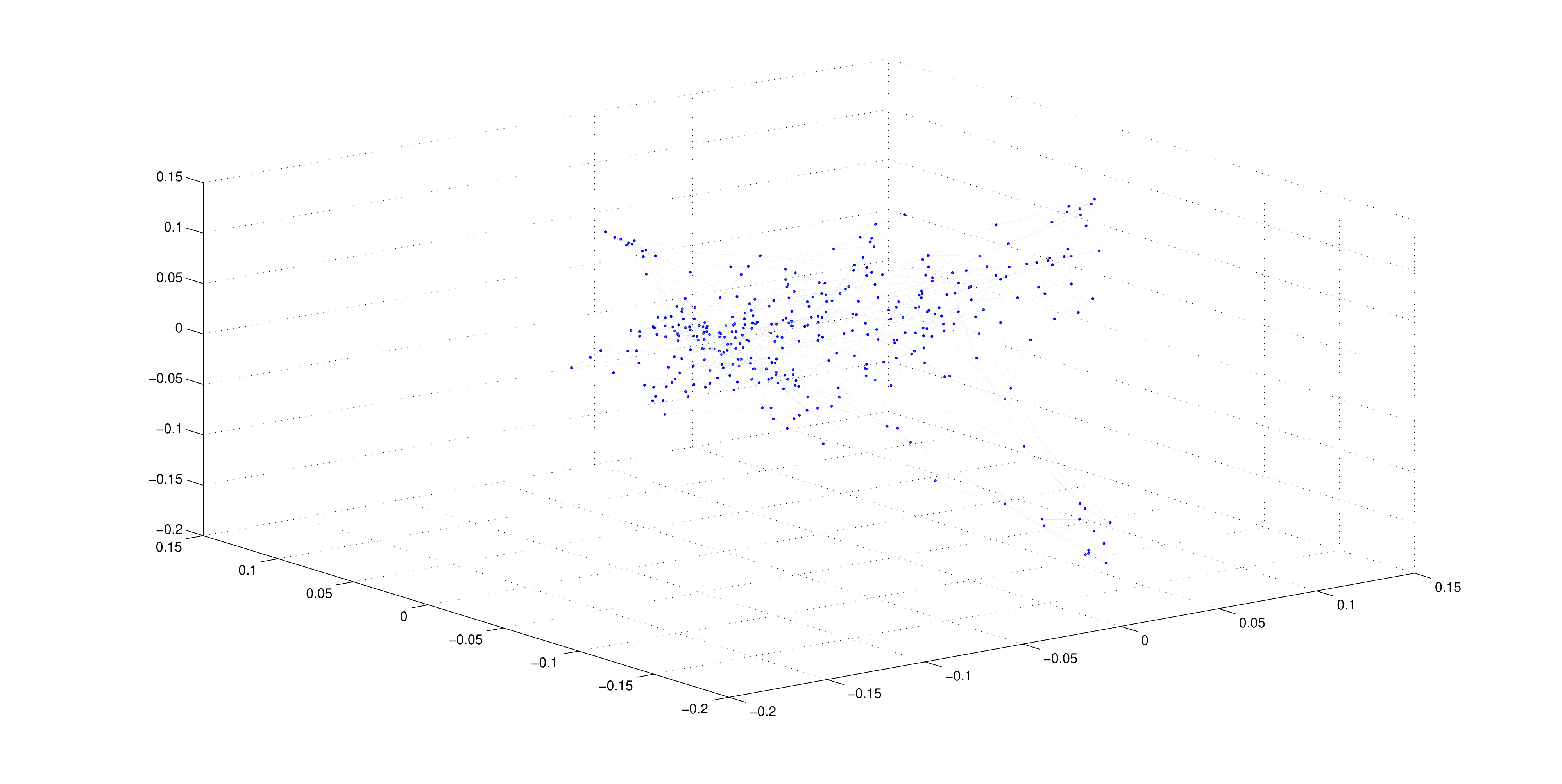}
\label{eigprojectionsgrewire3D}}
\label{eigprojection}
\caption[Optional caption for list of figures]{The Eigen-Projection Method}
\end{figure}
	
	The eigen-projection visualizations (Figure \ref{eigprojection}) allow us to make further distinctions between the \emph{C. elegans} brain and other networks.  In support of our previous observations, it is again clear that the nematode brain (Fig. \ref{eigprojectionworm} and \ref{eigprojectionworm3D} is not strictly fractal in structure.  The eigenfunction graphs of the Sierpinski gasket (Fig. \ref{eigprojectionsg} and \ref{eigprojectionsg3D}), once again display characteristics expected of self-similar fractals: a high degree of ordering and self-symmetry.  
	
	While, the eigenvalue counting function and Weyl ratios showed little distinction between the \emph{C. elegans} brain, and a random graph, these eigen-projections make a clear differentiation between the two.  The random graph (Fig. \ref{eigprojectionrandom} and \ref{eigprojectionrandom3D}) appears, as expected, more or less a scatter of points.  The \emph{C. elegans} brain, however, shows a definite structure with organized connectivity.  Thus while our previous results appeared more or less inconclusive, these eigen-projection techniques suggest that the structure of the \emph{C. elegans} neural network is \emph{not} a randomly connected system of neurons.  On the other hand, the \emph{C. elegans} neural network maintains its resemblance to a rewired Sierpinski gasket when plotted in eigenfunction coordinates (Fig. \ref{eigprojectionsgrewire} and \ref{eigprojectionsgrewire3D}).  While there is no effective way to quantify this heuristic similarity, it sustains its interest experimentally.  This additional similarity continues to suggest the presence of structural parallels.
	
	The eigen-projection method allows us to view not only the structural ordering, but also the functional organization of the \emph{C. elegans} neural network.  It is clear that the neurons are arranged roughly by neuron type.  There is a distinctive cluster of motor neurons (green), a larger sub-component of sensory neurons (red), and interneurons interspersed throughout the network (blue).  This indicates that although the brain may not posses the strict self-similarity of a fractal structure, it is indeed highly organized.  The neuron grouping evident in our visualization has most likely developed for entirely functional purposes: producing more efficient signal transfer and more organized communication between neurons.
	
\subsection{Small-World Network Properties}
\label{smallworld}

	When analyzing graphs as networks of nodes, it is useful to consider the nature of the graph's connections. To do this, we consider two functions defined on graphs: average clustering coefficient and average path length. The clustering coefficient, $c_{v}$, of a vertex $v$, is the probability that any two vertices neighboring $v$ are also connected to each other. Formally, this can be calculated by counting the number of edges between neighbors of $v$ and dividing by the number of total possible connections that could exist between neighbors of $v$. (For more details, see Methods \ref{methods3.4}). For a graph $G$, the average clustering coefficient, $c$, is the average of $c_{v}$ over all vertices.
	
	The path length between two vertices $u$ and $v$ is the shortest path along the graph's edges connecting $u$ and $v$ (Note that this path usually travels through a number of other vertices). Using Djisktra's algorithm, it is possible to rigorously determine the shortest path between a given vertex and each other vertex on the graph. By repeating the algorithm for each node on the graph, it is possible to determine the shortest path between each pair of vertices. It follows naturally that the average path length, $l$, is calculated by finding the arithmetic mean of the shortest paths between each pair of vertices on the graph. We use both $c$ and $l$ to analyze small-world behavior.
	
	One prominent theme in modern graph-theoretic and fractal research is the ``small-world phenomenon". This phenomenon is best described as the tendency for certain networks to have a much shorter path length than intuition suggests. For example, the term ``Six Degrees of Separation" suggests that any two people on earth are no more than six ``steps" away from each other, even though it seems that this number should be much larger than six. This phenomenon can be explained by the generalization that humans live in relatively large and tight-knit social networks, and thus have a relatively high $c$ value. However, each person also likely possesses a few long distance relationships, allowing $l$ to attain very small values without lowering the value of $c$ significantly. Thus, small-world networks are (generally) defined as networks which have a much higher $c$ value than random networks, but maintain a value of $l$ only slightly larger than that of a random network \cite{WS33}.

\begin{figure}[h!]{
\begin{center}
\begin{tabular}{ | l | c | r | }	
\hline
	Graph & Clustering Coefficient & Average Path Length\\ \hline
	Sierpinski Gasket, Level 5 & 0.4495 & 17.3721\\ \hline
	Random(Sierpinski Gasket) & 0.0104 & 5.748\\ \hline
	Sierpinski Gasket Rewire $p=0.15$ & 0.2843 & 7.3833\\ \hline
	Random(SG Rewire) & 0.0104 & 5.748\\ \hline
	\emph{C. elegans} Neural Network & 0.3371 & 2.5377\\ \hline
	Random(\emph{C. elegans} Neural Network) & 0.0581 & 2.3458\\
	\hline
	\end{tabular}
	\end{center}
	\label{c and l table}}
	\caption{Clustering Coefficient and Path Length}
	\end{figure}

	Small-world networks arise quite often in the natural sciences, as they allow for the efficient transfer of information while maintaining a certain level of complexity. There is a great deal of research which suggests that neural networks possess small-world properties \cite{BMAB31, SH32}. Our findings propose that \emph{C. elegans} is no exception.  As Figure \ref{c and l table} shows, the \emph{C. elegans} neural network has an average path length only slightly larger than that of its associated random network (see Methods \ref{methods3.5} for how these `associated random networks' were constructed). At the same time, the average clustering coefficient for \emph{C. elegans} is six times larger than that of its associated random network. This being the case, the neural network of \emph{C. elegans} satisfies the small-world properties as defined by \cite{WS33}.

	It should be noted that this discussion of small-world networks is relevant to our discussion of self-similar fractals. Certain research \cite{CS34} suggests that there is an apparent dichotomy between fractal structures and small-world networks. To illustrate this idea, refer again to Figure \ref{c and l table}.  It is apparent that the $5^{th}$ level Sierpinski Gasket is not small-world in nature: its average path length is significantly larger than that of its associated random network.  Before we continue our discussion of small-world networks, we must define the neighborhood of a graph.
	
\begin{figure}[!]
\centering
\subfigure[\emph{C. elegans} neural network]{
\includegraphics[width=.45\linewidth]{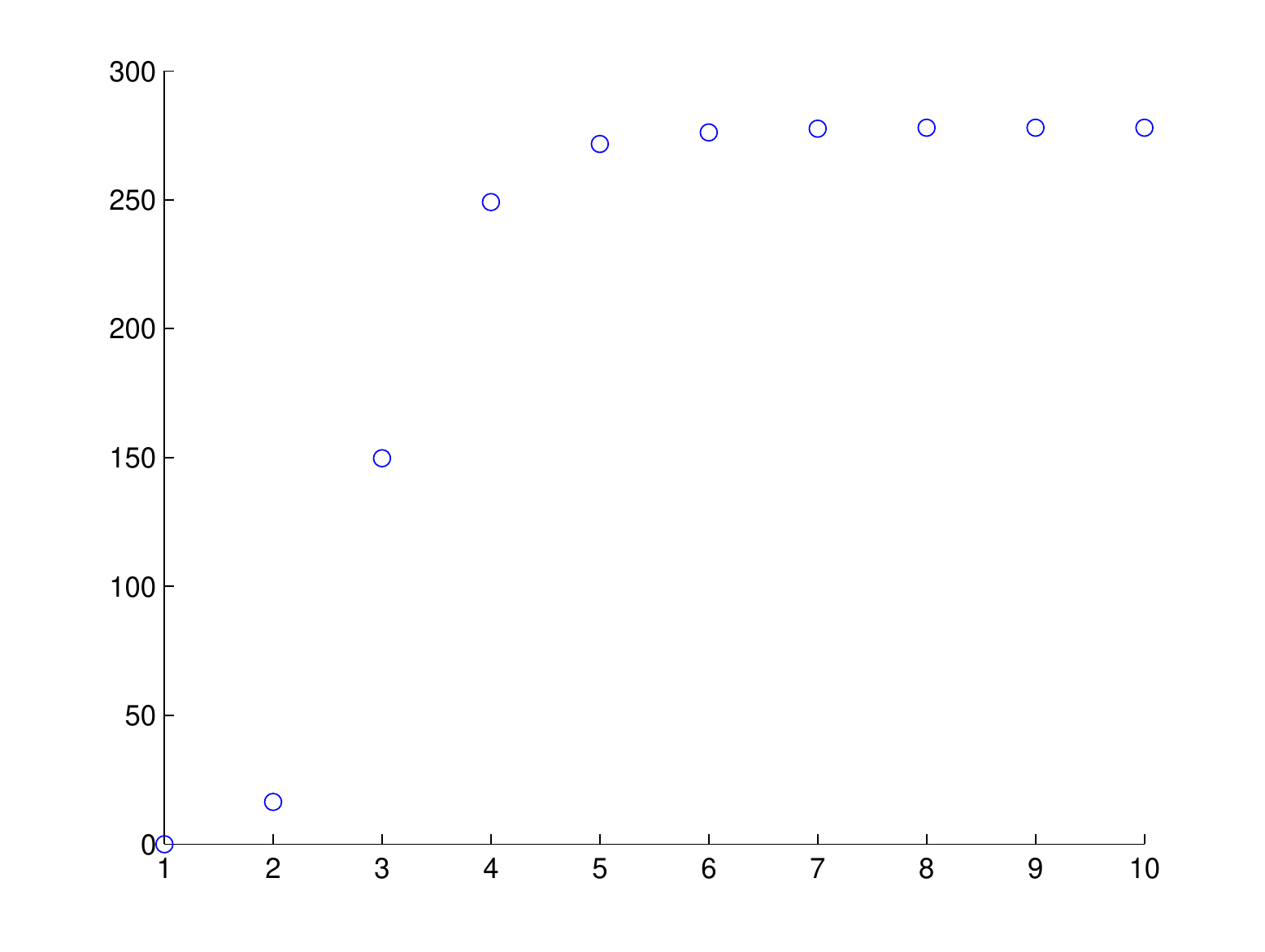}
\label{wormneighbors}}
\subfigure[\emph{C. elegans} neural network, log-log]{
\includegraphics[width=.45\linewidth]{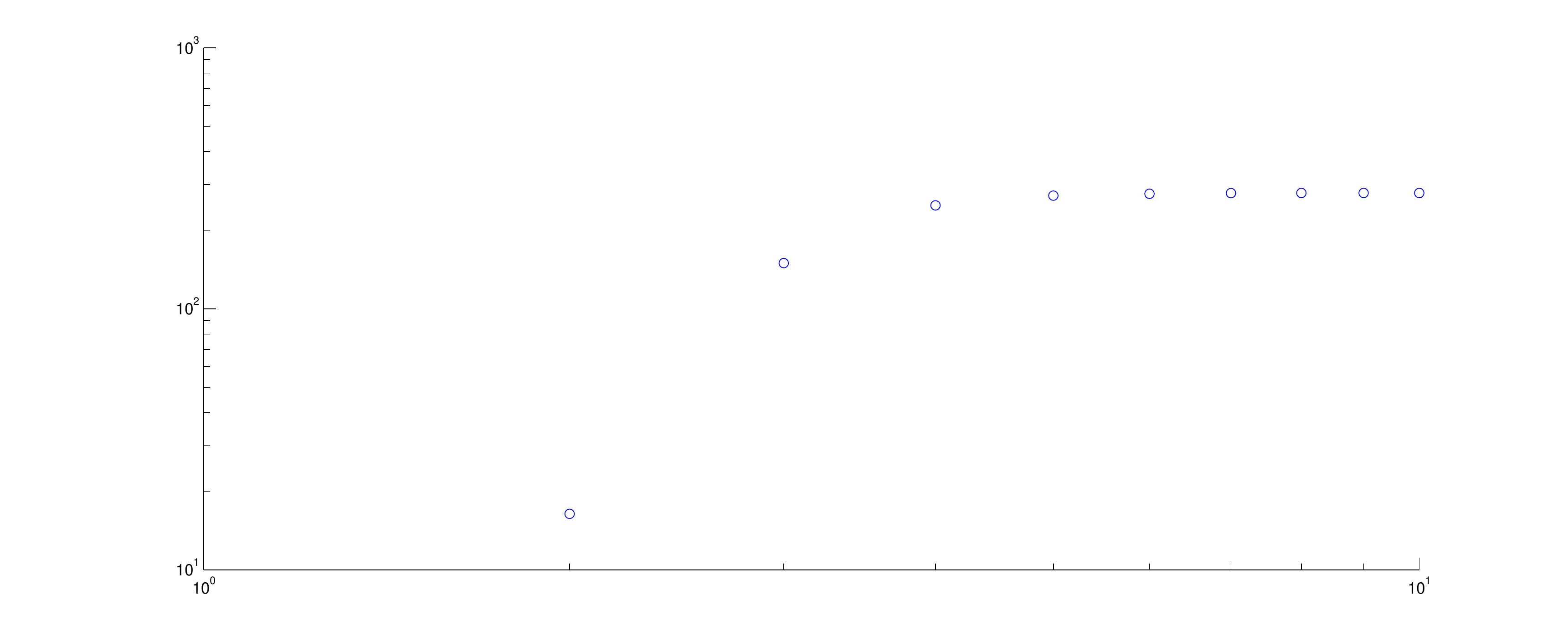}
\label{wormneighborsloglog}}
\subfigure[\emph{C. elegans} neural network, linear-log]{
\includegraphics[width=.45\linewidth]{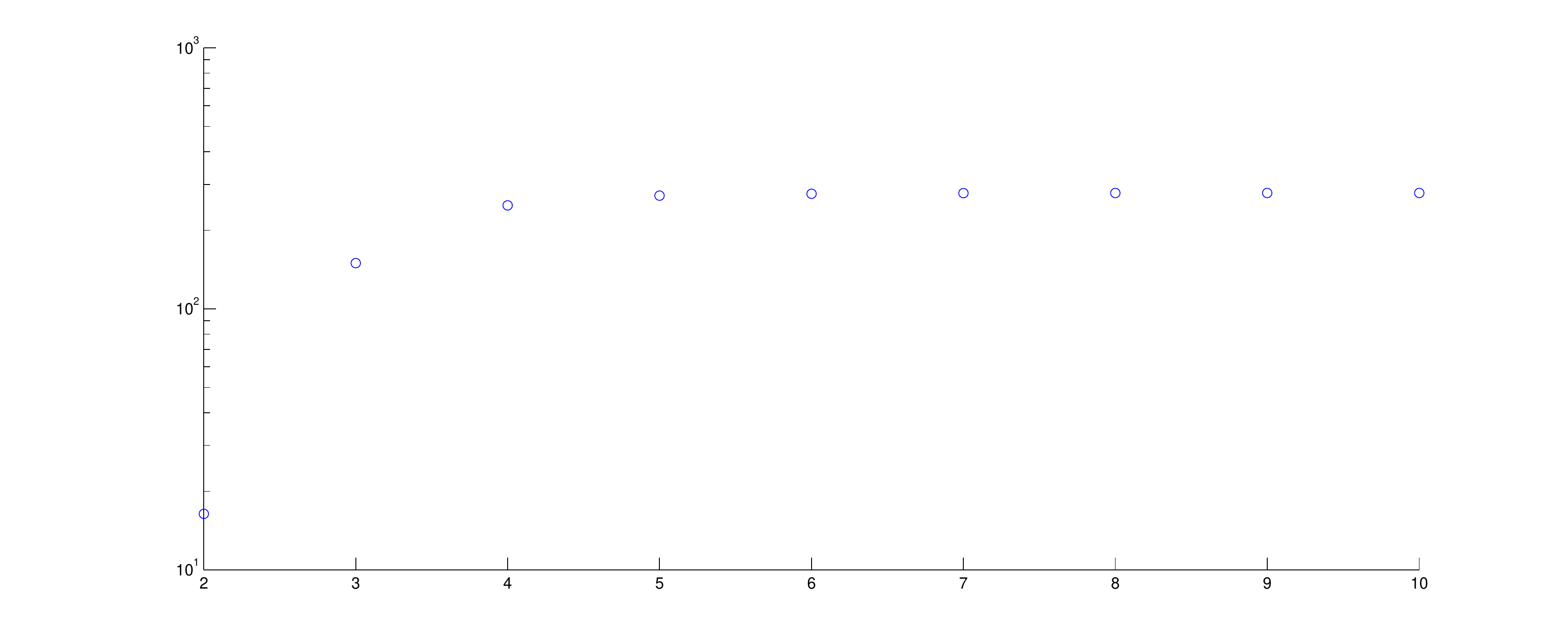}
\label{wormneighborslinearlog}}
\subfigure[Sierpinski Gasket, Level 5]{
\includegraphics[width=.45\linewidth]{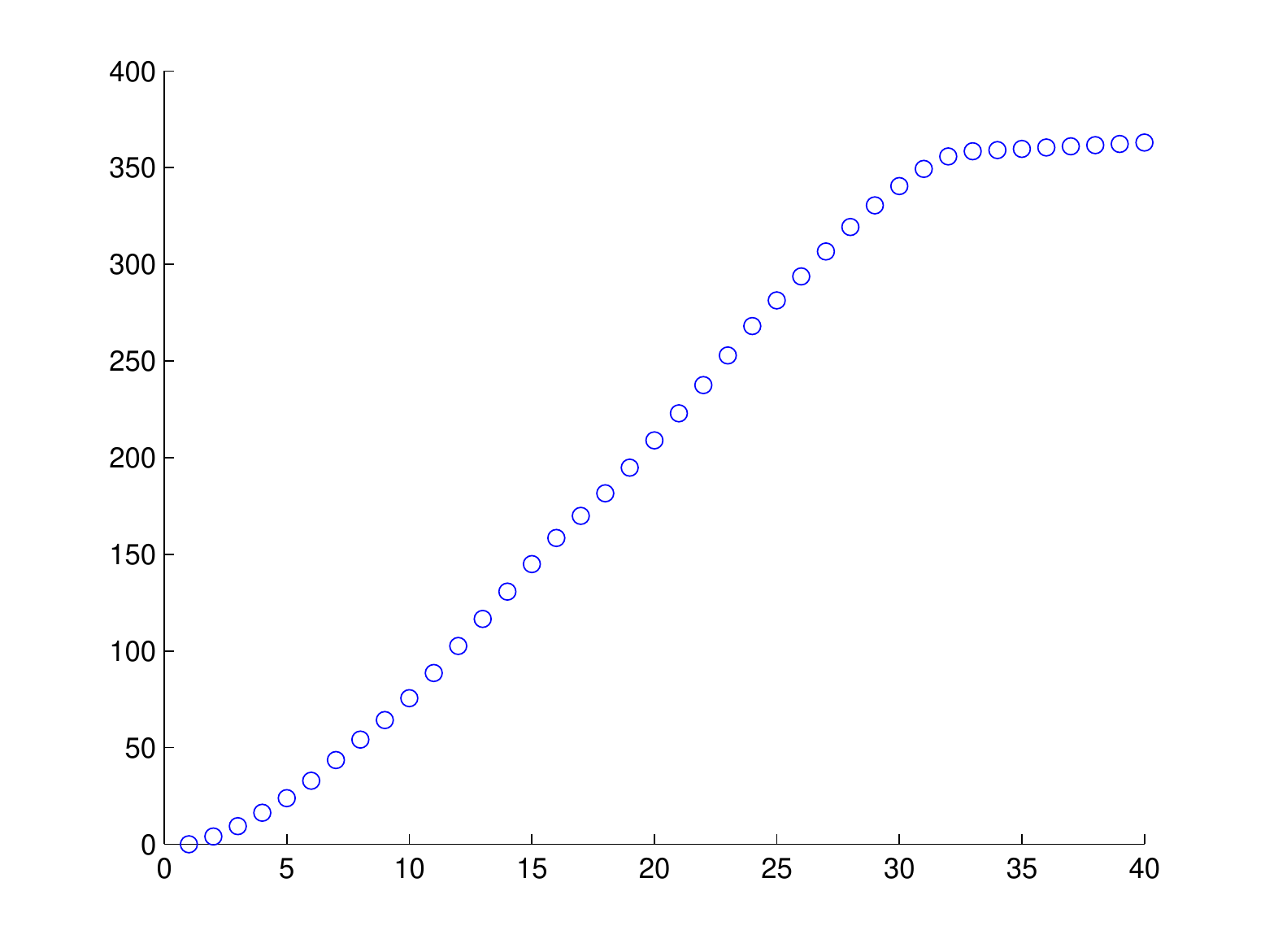}
\label{sierpinskineighbors}}
\label{neighbors}
\caption[Optional caption for list of figures]{Graph Neighborhoods}
\end{figure}

	The neighborhood of a graph, $H(m)$, where $m$ is a positive integer, is also useful in analyzing small-world networks. For our purposes a neighborhood of size $m$ around a vertex $v$ is the set of all vertices that can be reached in $m$ steps or less from $v$ (see Methods \ref{methods3.6} for a rigorous definition).  The behavior of $H(m)$ can tell us quite a bit about structure of a graph, $G$. If we look at the growth characteristics of $H(m)$ for small $m$, we begin to understand the localized neighborhoods of $G$. Certain research suggests that $H(m)$ grows exponentially for small $m$ if $G$ is a small-world network \cite{CS34}. Likewise, $H(m)$ tends to grow polynomially if $G$ is a self-similar fractal. Figure \ref{wormneighbors} shows a plot of $H(m)$ for the \emph{C. elegans} neural network.  Figures \ref{wormneighborsloglog} and \ref{wormneighborslinearlog} show the same plot on log-log axes and linear-log axes, respectively. Unfortunately, the number of relevant points is simply too small to make a concrete statement as to whether or not this $H(m)$ shows exponential or polynomial growth.  On the other hand, $H(m)$ is clearly polynomial for the Sierpinski Gasket (Fig. \ref{sierpinskineighbors}).  In fact, it appears roughly linear. 
	
\noindent

	Thus far, our research suggests that \emph{C. elegans} neural network does not appear to structurally resemble a tree, a random graph, or a self-similar fractal. However, it does appear to possess small-world properties. This realization motivated our work with network-rewiring, related to that done by Watts and Strogatz. In \cite{WS33} they showed that moving connections in an ordered network, with a certain probability $p$, led to some interesting changes in graph structure. Namely, when $p$ is small, a slight increase in $p$ causes a large drop in $l$ but does not change $c$ appreciably: thus the network takes on small-world characteristics.  Intuitively this can be explained by the fact that these sparse random connections don't change a graph's strong localized structure, but it is now easier to travel long distances via these new connections which can span large gaps.  
	
	This was our motivation in analyzing the rewired Sierpinski Gasket (seen throughout previous sections).  Self-similar fractals are highly ordered networks, as were the ordered graphs analyzed in \cite{WS33}.   By rewiring certain known fractal structures, we were able to observe if the resulting small-world network was comparable to our \emph{C. elegans} neural network, which has been demonstrated to have small-world properties (See Methods \ref{methods3.7} for our rewiring algorithm).  This would suggest that the \emph{C. elegans} brain has localized self-similarity with interspersed gap-spanning connections, allowing for more efficient signal transfer.  Figure \ref{c and l table} shows the results of rewiring the $5^{th}$ level Sierpinski Gasket. Note that the random graphs associated with the Sierpinski Gasket and its rewiring are identical because rewiring preserves the number of connections, and thus the average degree. It is clear that the rewired Sierpinski Gasket is an example of a small-world network based on our definitions, and thus maintains interest as a comparison to the \emph{C. elegans} neural network. While previous evaluations inferred similarities between these two networks, we now know that both show small-world properties as well.

\subsection{Energies and Spacial Variances}

Next, we analyzed energies and spacial variances on each of our graphs.  Using an eigenfunction of a graph's Laplacian, $\varphi$, one can calculate a graph energy specific to $\varphi$ (See Methods \ref{methods3.8}).  
Knowing the resistance between any two vertices and a constant $\gamma$, one can also use this $\varphi$ to calculate a spacial variance (See Methods \ref{methods3.9}).  These two quantities allow us to observe localization of eigenfunctions.  This localization occurs when eigenfunctions are approximately zero except for in a localized region.  Localized eigenfunctions are a feature of certain self-similar fractals, such as the Sierpinski Gasket  \cite{OSLT35}.  
\ \ \ \

The energies and spacial variances (at $\gamma = 1$) of the eigenfunctions of the \emph{unnormalized} Laplacian of the \emph{C.  elegans} neural network were compared to those of other graphs.  These included random graphs, fractal graphs, small-world networks, and random trees.  Distributions of these values were plotted for each network.  Any resemblance in the distributions of the energies and spacial variances could indicate some similarity in the structure of the graphs \cite{OSLT35}.
\ \ \ \

\noindent
\begin{figure}[!]
\centering
\subfigure[\emph{C. elegans} neural network energies]{
\includegraphics[width=.45\linewidth]{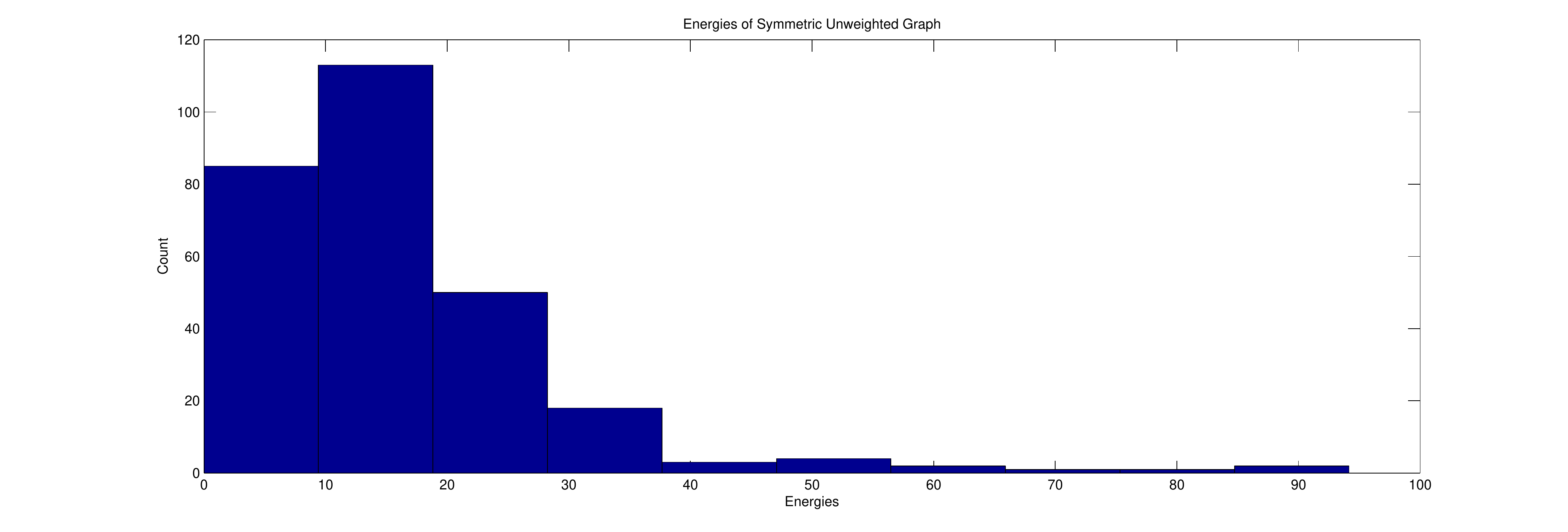}
\label{wormenergy}}
\subfigure[\emph{C. elegans} neural network spacial variances)]{
\includegraphics[width=.45\linewidth]{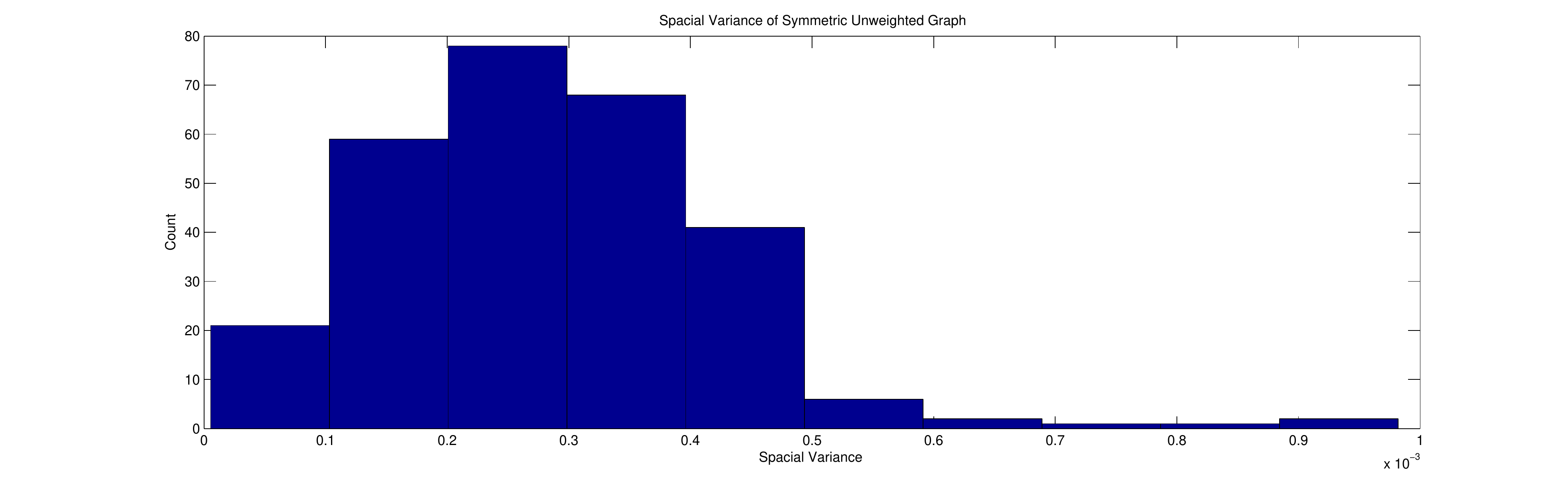}
\label{wormvariance}}
\subfigure[Random graph ($n=279, p=0.07$) energies]{
\includegraphics[width=.45\linewidth]{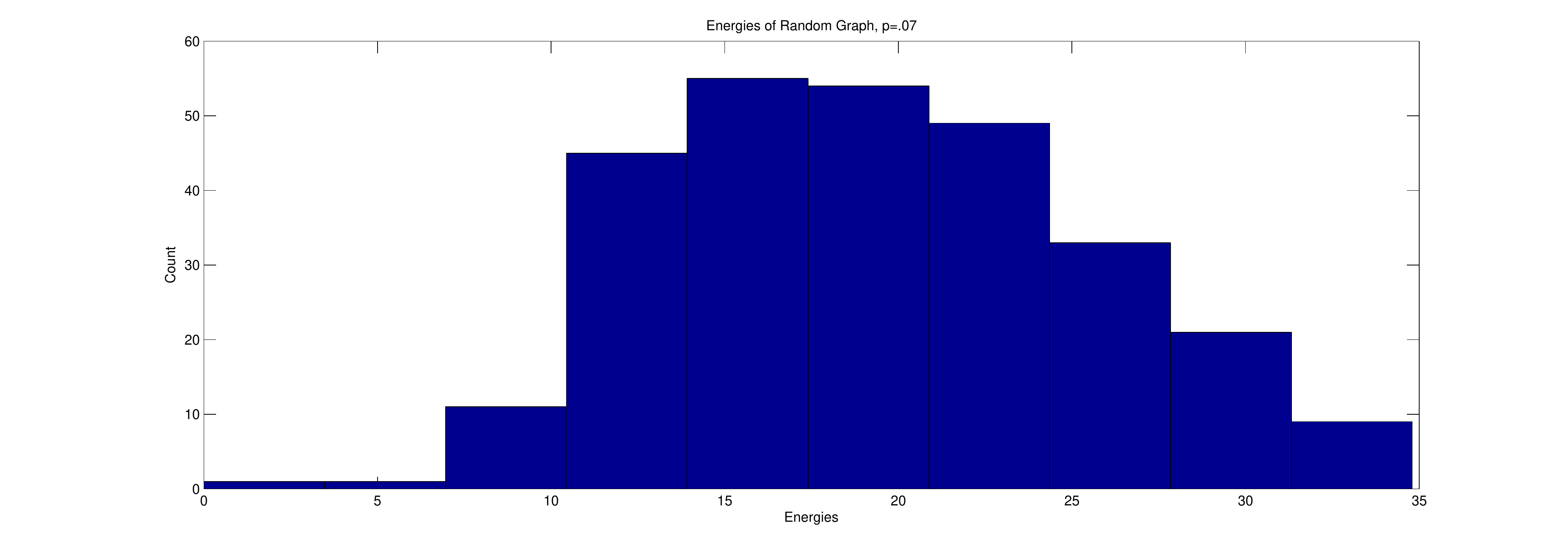}
\label{randomgraphenergy}}
\subfigure[Random graph ($n=279, p=0.07$) spacial variances]{
\includegraphics[width=.45\linewidth]{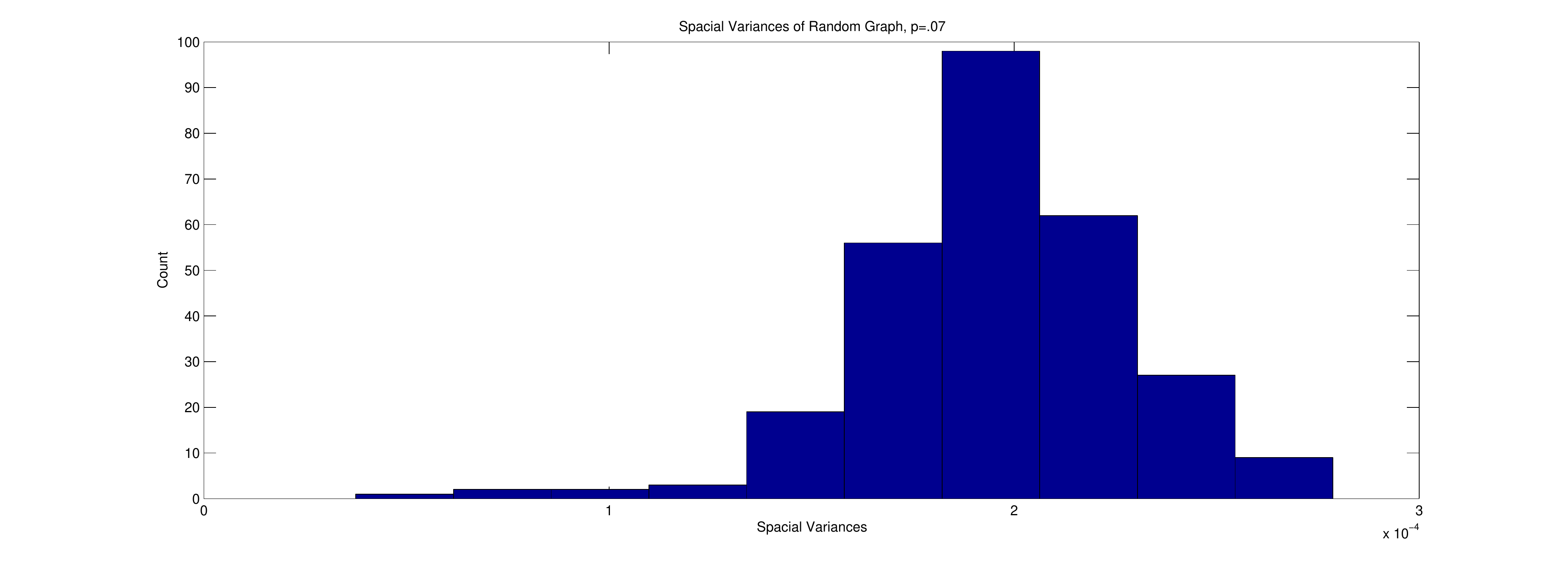}
\label{randomgraphvariance}}
\subfigure[Sierpinski Gasket, Level 5 energies]{
\includegraphics[width=.45\linewidth]{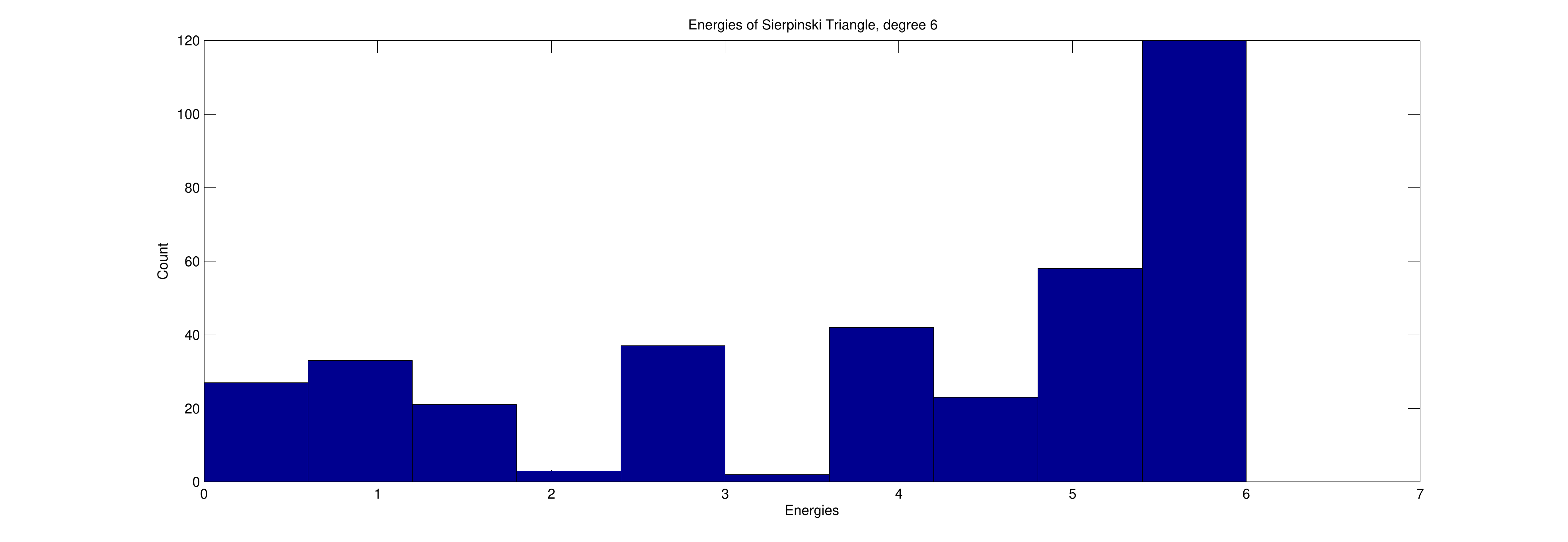}
\label{sierpinskigasketenergy}}
\subfigure[Sierpinski Gasket, Level 5 spacial variances]{
\includegraphics[width=.45\linewidth]{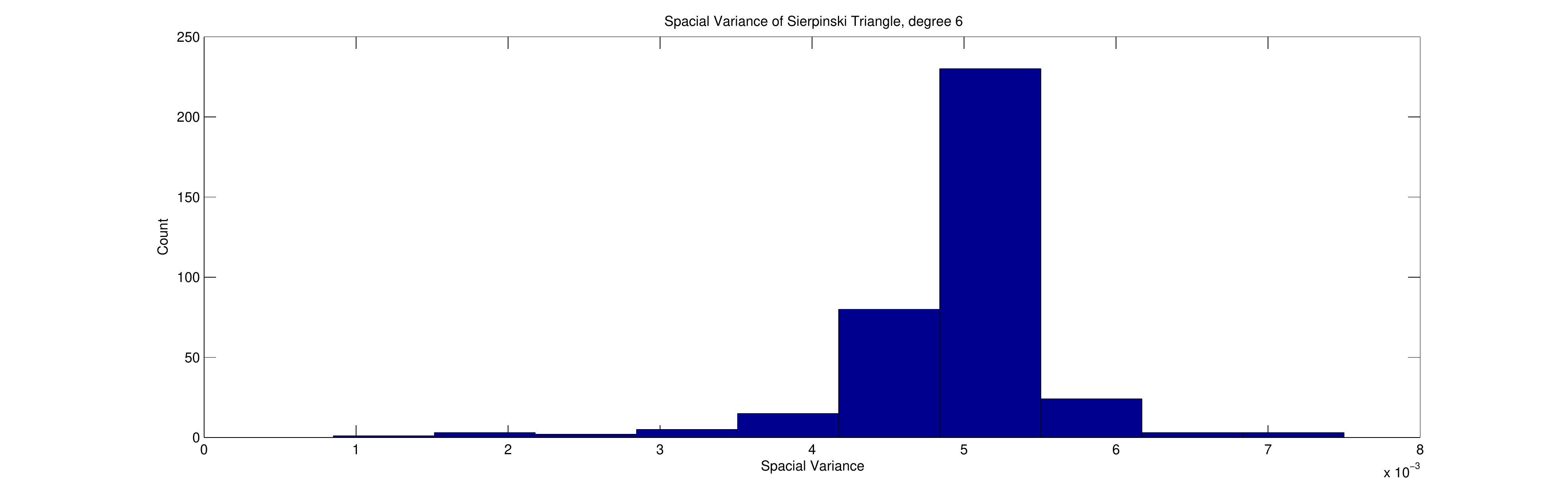}
\label{sierpinskigasketvariance}}
\subfigure[Random Tree ($n=279, m=10$) energies]{
\includegraphics[width=.45\linewidth]{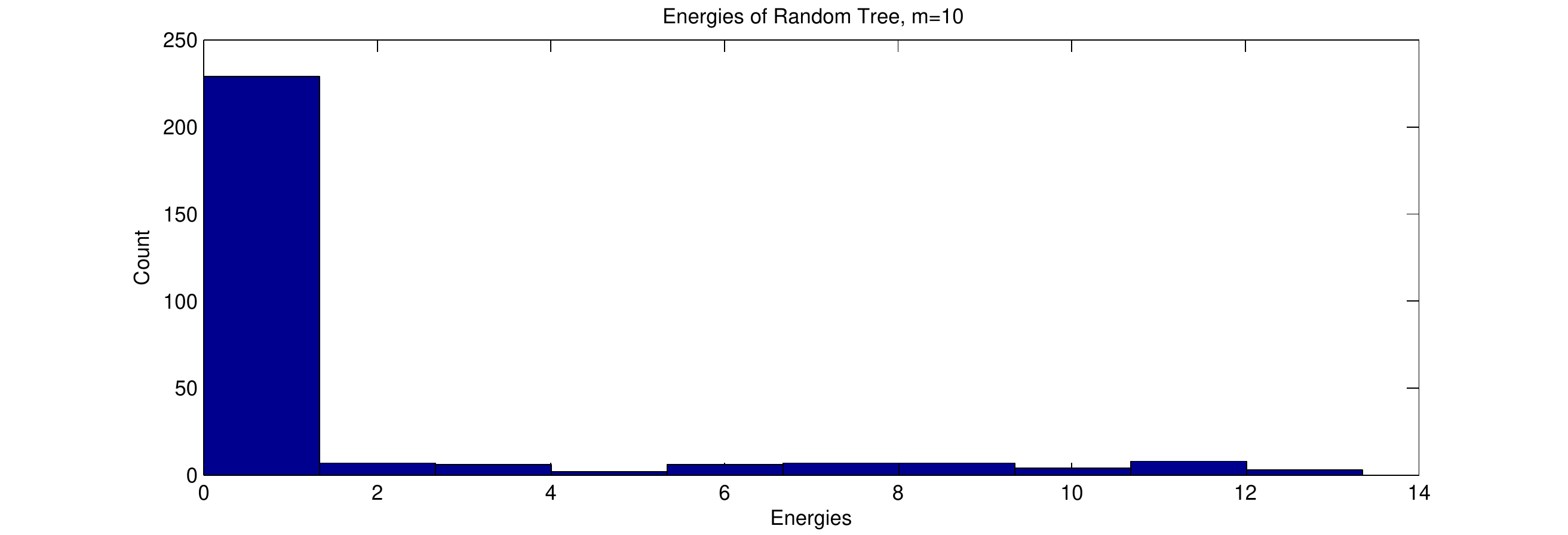}
\label{randomtreeenergy}}
\subfigure[Random Tree ($n=279, m=10$) spacial variances]{
\includegraphics[width=.45\linewidth]{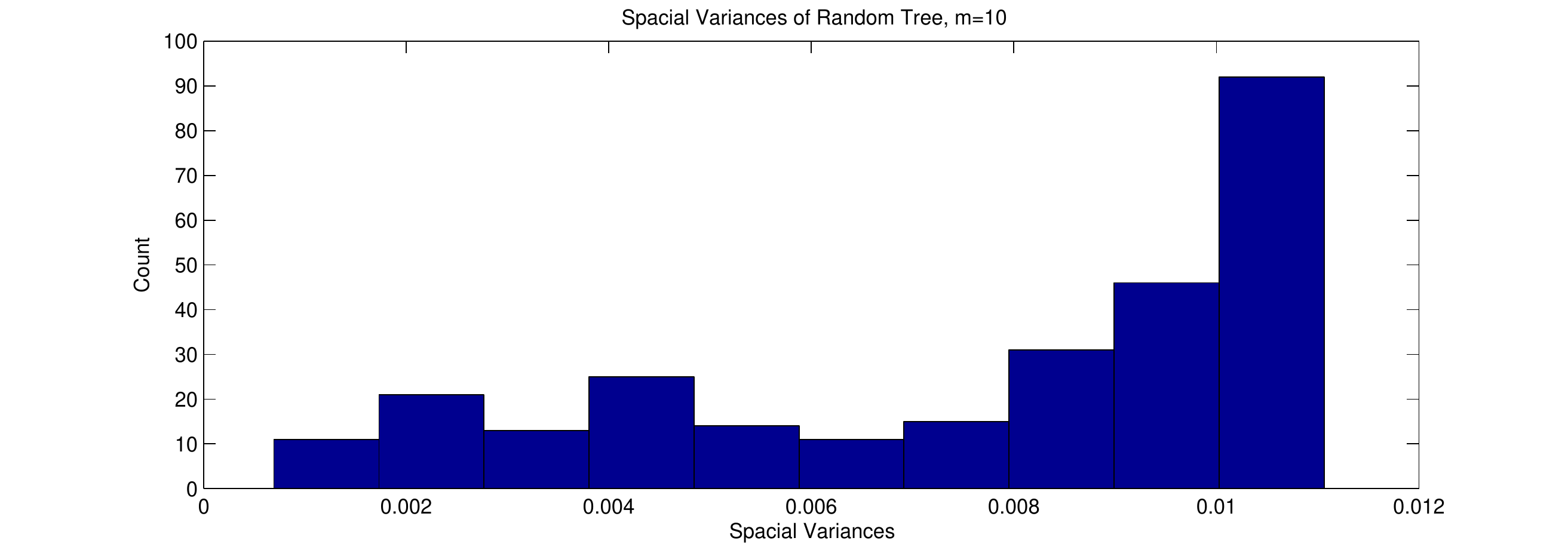}
\label{randomtreevariance}}
\subfigure[Sierpinski Gasket Rewiring ($p=0.15$), energies]{
\includegraphics[width=.45\linewidth]{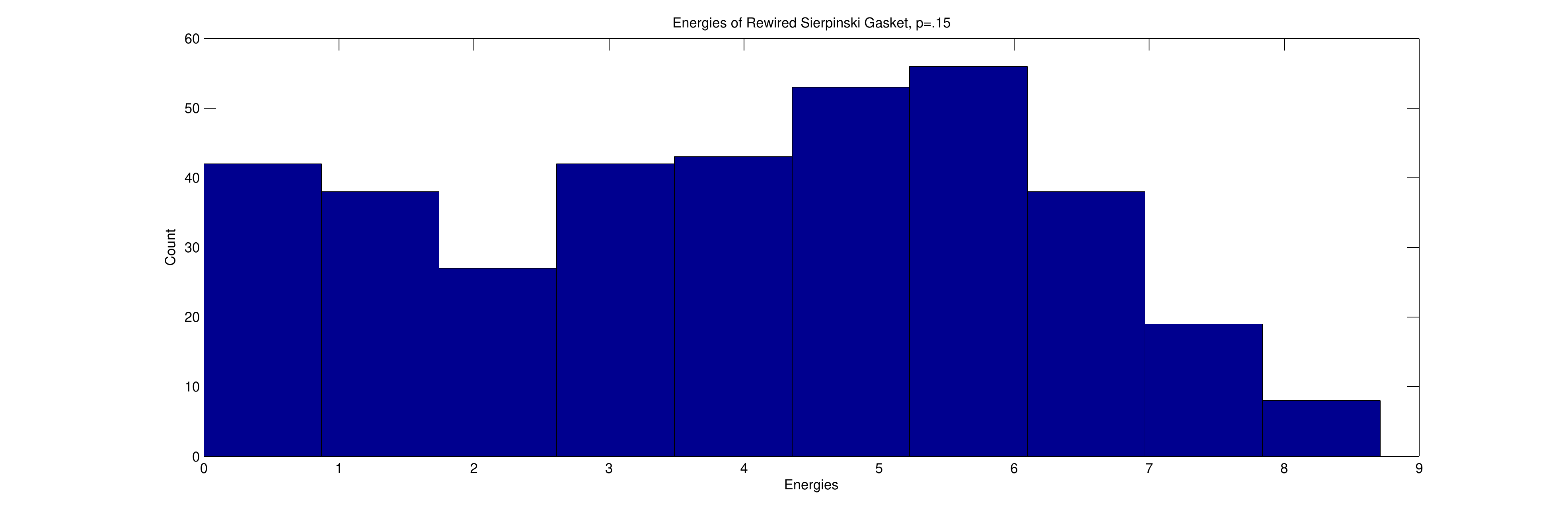}
\label{rewiresierpinskigasketenergy}}
\subfigure[Sierpinski Gasket Rewiring ($p=0.15$), spacial variances]{
\includegraphics[width=.45\linewidth]{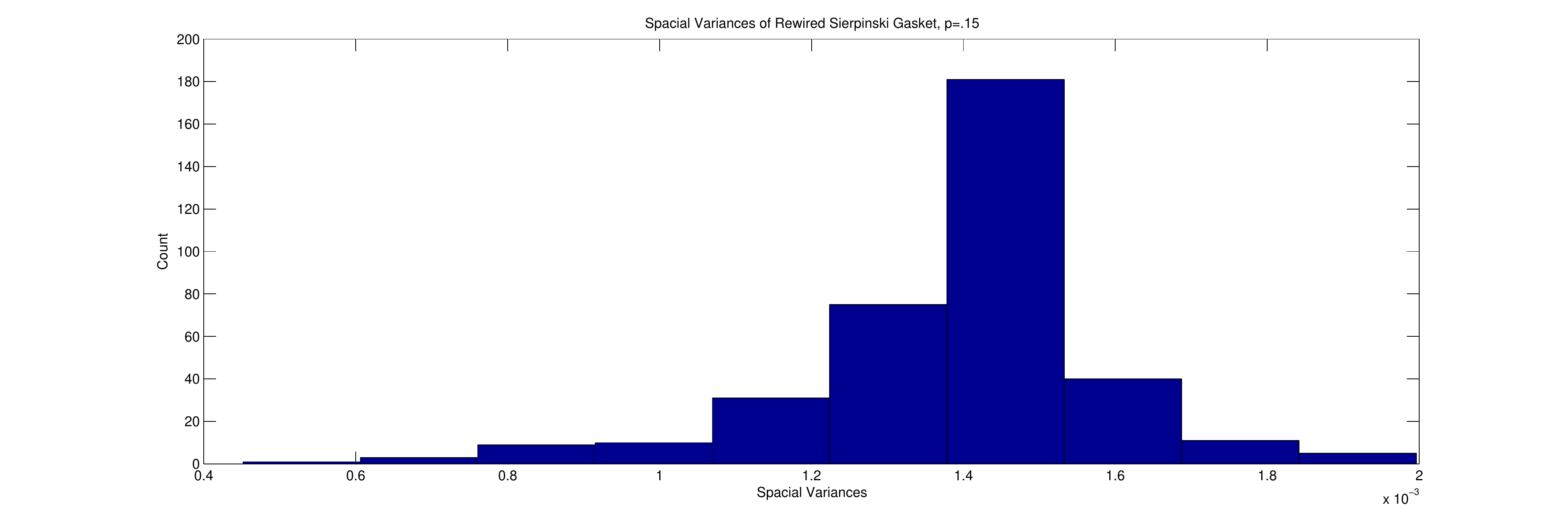}
\label{rewiresierpinskigasketvariance}}

\label{energyandspacialvariance}
\caption[Optional caption for list of figures]{Graph Energies and Spacial Variances}
\end{figure}

The energies and spacial variances of the random graph (Fig. \ref{randomgraphenergy} and \ref{randomgraphvariance}) were quite different from those of the neural network.  In the random network, the distributions of these values tend to be peaked and symmetric, whereas those of the neural network are more spread-out and skewed (Fig. \ref{wormenergy} and \ref{wormvariance}). This evidence suggests that the neural network of \emph{C. elegans} is not randomly distributed.  Intuitively, this is what one would expect, and is consistent with our earlier observations.
\ \ \ \

Next, we continued our comparison of the neural network of \emph{C. elegans} to known self-similar fractals, once again using the $5^{th}$ subdivision of the Sierpinski Gasket.  The energies and spacial variances of this fractal (Fig. \ref{sierpinskigasketenergy} and \ref{sierpinskigasketvariance}) vary considerably from those of the neural network.  Although the spacial variances of all graphs were roughly symmetric, the variances of the neural network were much lower than those of the Sierpinski Gasket.  On the neural network, the spacial variances were all less that $1 \times10^{-3}$, whereas those of the fractal network ranged between $1\times10^{-3}$ and $8\times10^{-3}$.  Furthermore, the distribution of energies of the neural network was skewed right and ranged from 0 to almost 100, while the energies of the Sierpinski Gasket were very jagged (no resemblance to a normal distribution) and contained many gaps.  In addition, all of these energies were in a narrow range between 0 and 6.   These differences are in agreement with earlier results, suggesting that the nematode brain is not strictly fractal in structure.  
\ \ \ \

We also compared the \emph{C. elegans} neural network to a random branching tree (Fig \ref{randomtreeenergy} and \ref{randomtreevariance}).  The energies and spacial variances of the random tree are again different from those of the neural network.  The energy distribution of the tree was peaked, with most values falling between 0 and 10, whereas those of the neural network were distributed over a wider range, from 0 to 100.  The energies of the neural network are skewed right.  The differences in spacial variances were even greater.  The spacial variances of the neural network lie between 0 and $1\times10^{-3}$, whereas those of the random trees are spread over a larger range, up to about $1.5\times10^{-2}$.  The distribution of the neural network is again skewed right, whereas that of the random tree is very much skewed left.  Once again, the neural network does not exhibit the same structural properties as a random-branching tree.  
\ \ \ \

Thus far, the ``small-world" rewired Sierpinski gasket has shown the most similarities to the \emph{C. elegans} neural network.   Once again we analyzed the rewiring of the fifth level Sierpinski triangle, with $p=0.15$.  This probability was chosen to generate a small-world network without disturbing much of the self-similar structure.  After rewiring, the energies of the graph (Fig. \ref{rewiresierpinskigasketenergy}) were distributed in nearly the same pattern as those of the original self-similar fractal, only the values were shifted slightly to the right.  The same is true of the spacial variances (Fig. \ref{rewiresierpinskigasketvariance}), except that this distribution was shifted to the left.  Although the range of the distributions on the rewired fractal is closer to that of the neural network, the distributions themselves remain markedly different from those of the \emph{C. elegans} brain.  This leads us to believe that the structure of the \emph{C. elegans} neural network, although small-world in nature, does not structurally resemble the Sierpinski Gasket.  On the other hand, the spacial variances of the \emph{C. elegans} neural network are considerably lower in magnitude than those of the Sierpinski Gasket.  This suggests localization of eigenfunctions, which could in turn indicate the presence of self-similarity in the network.  
 \ \ \ \ 

\subsection{Conclusions}
Using a variety of mathematical techniques including the eigenvalue counting function, Weyl ratios, and the eigen-projection method, analyzing small-world properties, and calculating of graph energies and spacial variances, we were able to uncover some structural properties of the \emph{C. elegans} neural network.  Although much previous research has been dedicated to applying fractal theory to neuroscience, our results suggest otherwise.  The known structure of the \emph{C. elegans} neural network does not exhibit properties characteristic of strictly self-similar fractal networks.  While some evidence suggests that there may be localized instances of self-similarity, we cannot quantify or definitively state such a conclusion.  Our research found additional structures which the \emph{C. elegans} neural network does not resemble: it does not exhibit the branching properties of a tree, nor does it constitute a randomly connected network. 

As to what we \emph{can} conclude, our research is consistent with related work, showing that the brain exhibits small-world network properties.  Furthermore, the network has highly localized eigenfunctions, which could suggest the presence of self-similar structures.  Further research would be required to determine the nature of these localized eigenfunctions.  Although the \emph{C. elegans} neural network does not appear to be random, tree-like, or fractal in structure, it is certainly highly ordered.  This organization most likely aids in functional efficiency of the system.  Further research is needed to determine a more refined view of the brain's structural properties.  Although \emph{C. elegans} has proven to be a useful model organism, with a well-defined map of its neural network, this network consists of only 279 nodes.  While this makes the system fairly efficient to study computationally, this small number of nodes makes network-analysis somewhat limited and rather unrevealing.  However, due to the difficulty in determining the exact layout of each neuron in the network, very few consistent complete-brain maps exist at this time.  Similar analysis as that provided in this paper, applied to a more complex or higher-order neural network, could potentially show more conclusive results.

%% file: Methods2.tex
\section{Methods}\label{methods}
In order to analyze only the framework of the \emph{C. elegans} neural network, we constructed a Laplacian matrix derived form the adjacency matrices in \cite{PVC30}.  We wanted to look only at connections between neurons, regardless of type or direction.  The network of chemical synapses sends signals in one direction only, resulting in a non-symmetric adjacency matrix, $C$.  In order to disregard this directionality, we added this matrix to its own transpose, $C'$, creating a symmetric matrix indexing all chemical connections.  We added this matrix to the adjacency matrix of the gap junction system, $G$ (which is already symmetric because these connections are bidirectional).  

\[
B=[b_{i,j}]= (C+C')+G  \text{ such that } i,j \leq 279
\]

We then normalized all non-zero entries of this combined matrix, $B$, to be 1 (in order to avoid multiplicity of connection), resulting in an adjacency matrix representing only the framework of the entire network, $A$.  

\[
A=[a_{i,j}]  \text{ where }  a_{i,j}=b_{i,j}/b_{i,j}=1 \text{ if } b_{i,j}>0, \text{ otherwise }  a_{i,j}=0 (\text{ when } b_{i,j}=0)
\]

It is then simple to produce a Laplacian matrix, $L$, as shown below:

\[
d_{j}=\sum_{i=1}^{i=279} a_{i,j} \text{ for each } j \leq 279
\]Note that $d_{j}$ is the degree of each vertex $j$.  The degree matrix, $D$, is now defined as:

\[
D=[d_{i,j}] \text{ where } d_{i,j}=d_{j} \text{ when } i=j \text{ otherwise } d_{i,j}=0 (\text{when } i \neq j)
\]

Then the Laplaican matrix, $L$, is given by:

\[
L=D-A
\]

\subsection{Generating Random Graphs and Trees}
\label{methodsrandom}

Throughout this project, we compared the \emph{C. elegans} neural network to both random graphs and branching trees.  In order to generate a Laplacian matrix representation of the random graphs, we used the following algorithm:

First, we fixed the number of vertices, $n$, and the probability of connection, $p$.  We then constructed an empty $n \times n$ matrix, $R= [r_{i,j}]$.

For each $r_{i,j}$ such that $i < j$ we assign a random value $a_{i,j}$ such that $0 \leq a_{i,j} \leq 1$ for all $i,j \leq n$.  If $a_{i,j} \leq p$ then $r_{i,j}=1$, otherwise $r_{i,j}=0$.

To produce an adjacency matrix of this graph, $A$, we must add $R$ to its own transpose:
$$A = R + R' $$
Using this adjacency matrix we can construct a Laplacian matrix using the method described previously.\\~\\

The algorithm for producing the Laplacian matrix of a random-branching tree is more involved.  Again, we fix the number of vertices, $n$, and also specify the maximum number of "children" from any given branch-point, $m$.  We create an empty $n \times n$ matrix, $T=[t_{i,j}]$

We begin by generating a random integer $a_1$ such that $0 < a_1 \leq m$, and take $t_{1,1}=a_1$.  This corresponds to the first vertex having $|a_1|$ branches.  To represent these branches in the matrix, we take $t_{1,j}= -1$ for $j=2, \cdots, a_1+1$ and $t_{i,1}= -1$ for $i=2, \cdots, a_1+1$.  

Next we move to all subsequent vertices.  Because no ``looping" exists in the structure of the tree, each node can only be connected to its parent vertex and its "children" vertices.  We take $S=\{j: t_{i,j}=0 \text{ for all } i \leq n\}$  Then $k$, where $k=$min($S$) is the smallest-labeled node which does not have a parent vertex, i.e. the first column with all 0 entries corresponds to the first point not yet connected. (Note in the case of vertex 2, $k=a_1+2$).  This vertex $k$ is the first "offspring" from the next branch-point.

Now, as above, for each remaining vertex $v$ we choose another random integer, $a_v$, such that $0 < a_v \leq \text{ min}(m, n-k+1)$ and take $t_{v,v}= a_v+1$.  (Note that vertex $v$ has $a_v$ children, however $a_v+1$ is the degree of node $v$, taking into account its parent-connection).  To represent the "offspring" branches of this vertex $v$, we use the following formula:

$$ t_{i,v}= -1\text{ for }i=k,k+1,\cdots, k+(a_v-1)$$
and
$$ t_{v,j}= -1\text{ for }j=k,k+1,\cdots, k+(a_v-1)$$

We use min$(m, n-k+1)$ when choosing $a_v$ to avoid adding more vertices than the $n$ which we originally fixed.  This algorithm, when repeated for each vertex $v$, produces the Laplacian matrix of a random branching tree.

\subsection{The Eigenvalue Counting Function and Weyl Ratios}
\label{methods3.1}

	For a given graph Laplacian matrix, $L$, the eigenvalue counting function, $N(x)$ is a cumulative frequency function on the spectrum of the matrix where:
\[
N(x) = \#\{\lambda_{j} \leq x\} \text{ where each } \lambda_{j}  \text{ is an eigenvalue of } L
\]
The growth of $N(x)$ is approximately $x^\alpha$, thus the relevant portion of each graph, when using a logarithmic scale, appears to be linear.  A line of best fit was found for each relevant interval, and the slope, $\alpha$, calculated.  Using this $\alpha$, we plotted the Weyl ratio, $W(x)$, such that:
\[
W(x)= N(x)/x^{\alpha}.
\]
These Weyl ratios allow us to examine the spectrum of each matrix, looking for elements such as symmetry and periodicity. \cite{BHS26}

\subsection{Normalizing a Laplacian Matrix}
\label{methods3.2}

	We used two different forms of the graph-Laplacian matrix: the standard Laplacian matrix and the degree-normalized Laplacian.  In the case of eigen-projections, we utilize the degree-normalized matrix.  We define the degree matrix, $D$, as before: a diagonal matrix whose non-diagonal elements are 0, and each entry $d_{j,j}$ is the degree of the $j^{th}$ vertex. Using both the standard Laplacian, $L$ and its corresponding degree matrix, $D$, we produce the degree-normalized Laplacian, $Q$:
\[
Q = D^{-1/2} L D^{-1/2}
\]

\noindent
\begin{figure}[h!]
\centering
\subfigure[Un-normalized Laplacian, $(\varphi_2,\varphi_3)$]{
\includegraphics[width=.45\linewidth]{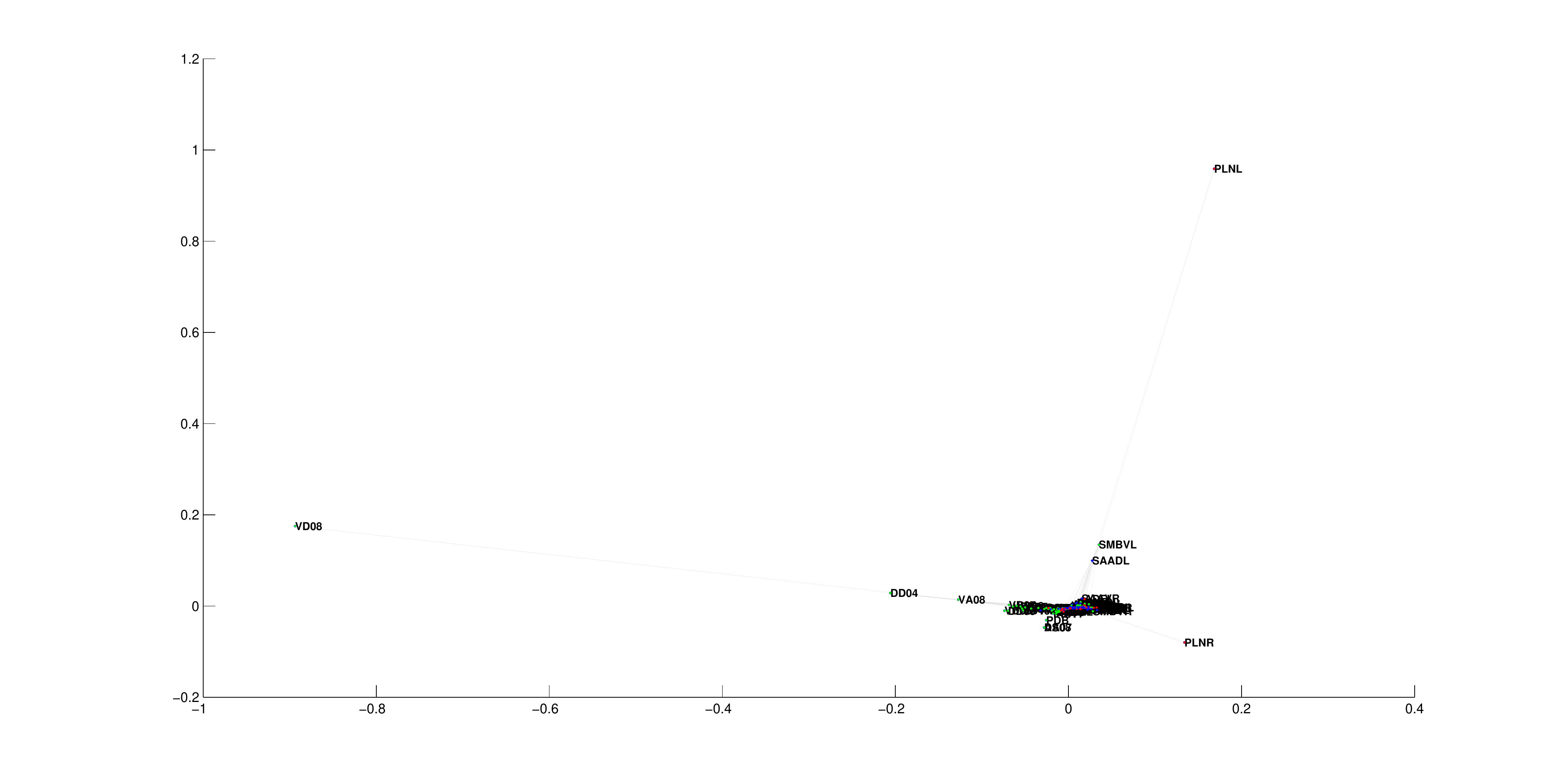}
\label{eigprojectionwormunnorm}}
\subfigure[Normalized Laplacian $(\varphi_2,\varphi_3)$]{
\includegraphics[width=.45\linewidth]{EigProjectionWorm.pdf}
\label{eigprojectionwormnorm}}

\label{normalize}
\caption[Optional caption for list of figures]{\emph{C. elegans} neural network}
\end{figure}

Using the degree-normalized Laplacian has many aesthetic advantages, as shown in Figure \ref{eigprojectionwormunnorm} and \ref{eigprojectionwormnorm}.  The normalized matrix also has all eigenvalues $\lambda_j$ such that $0 \leq \lambda_j \leq 2$.

\subsection{Graphing in Eigenfunction Coordinates}
\label{methods3.3}
	We found all eigenvalues, $\lambda_k$, and their corresponding eigenfunctions, $\varphi_k$, for each matrix.  Given two eigenfunctions $\varphi_i$ and $\varphi_j$, (such that $i \neq j$) we then plotted the ordered pair $(\varphi_i(n), \varphi_j(n))$ for each $n$ from 1 to 279, as described in \cite{BKNPPT}.  The first eigenvalue of any Laplacian matrix is always 0, corresponding to a constant eigenfunction.  Thus we only consider $\varphi_i$ and $\varphi_j$ with $i ,j \geq 2$.  Edges were then added between points to represent relevant connections, and the same color-coding as \cite{VPC25} was used: where red represents sensory neurons, green are motor neurons, and blue indicates interneurons.  The same process was then repeated in three dimensions, plotting $(\varphi_i(n), \varphi_j(n), \varphi_k(n))$ for some $i, j, k \geq 2$, such that $i \neq j \neq k $.

\subsection{Clustering Coefficient}
\label{methods3.4}

	The clustering coefficient is a common measure for vertices on a graph. It is typically measured on graphs with unweighted edges.  For a graph $G$ and a given vertex $v$, let $e_v$ denote the number of connections that exist between the neighbors of $v$.  Take $d_v$ as the number of neighbors of $v$ (the degree of vertex $v$). Then the clustering coefficient of vertex $v$, $c_v$, is given by:
\[
c_{v} =\frac{2e_v}{d_v (d_v -1)}
\]
Note that total number of possible connections among neighbors of $v $ is $\frac{d_v(d_v - 1)}{2}$.  

	Therefore, the clustering coefficient is essentially the probability that two neighbors of $v$ are connected. For a graph $G$ with $n$ vertices, the average clustering coefficient, $c$, is defined as:
\[
c=\frac{1}{n}\sum_{v=1}^{n} c_v
\]

\subsection{Generating a Related Random Graph for Small-World Analysis} 
\label{methods3.5}

	In order to analyze our networks for small-world properties, it was useful to compare these graphs to those of similar networks with randomly assigned edges.  Small-world networks are nearly as well-connected as random graphs, but possess a surprisingly well-localized structure.  We developed the following algorithm for this process:
		
		For a graph $G$ with $n$ vertices, let $k$ be the number of edges on $G$.  Therefore the average number of edges per vertex is $k/n$.  Furthermore, the probability that any two random vertices are connected, $p$, is given by the number of existing connections divided by the total possible connections: 
\[		
p= \frac{k}{\frac{n(n-1)}{2}} = \frac{2k}{n(n-1)}
\]
Next we generate a random graph, $Rand(G)$, with $n$ vertices and a $p$ probability of connection between two vertices (See Methods \ref{methodsrandom}).   We then compute $c$ and $l$ for $G$ and $Rand(G)$.\\~\\ 
$G$ has small-world network properties [33] if: 
 
 1. $l (G) \gtrsim l (Rand(G))$
 
 and
 
 2. $c (G)$ > > $c (Rand(G))$ where $c (Rand(G)) \approx p$ 

\subsection{The Neighborhood of a Graph}
\label{methods3.6}

	On a graph $G$ with $n$ vertices, the neighborhood, $H(m)$ where $m$ is a positive integer, is useful in analyzing small-world networks. A neighborhood of size $m$ around  vertex $v$ is the set of all vertices that can be reached from $v$ in $m$ steps or less. We shall refer to the number of vertices reachable in $m$ steps or less as $H_{v}(m)$. To get a sense of the global neighborhood size on the graph, we can average $H_{v}(m)$ over each vertex $v$ in $G$.  This gives us a global $H(m)$ for a given $m$:
\[	
H(m)= \frac{1}{n}\sum_{v=1}^{n} H_{v}(m)
\]
 It is clear that $H(m)$ is non-decreasing and as long as $G$ is complete, $H(m)$ achieves its maximum, $n$, for finite $m$.

\subsection{Graph Rewiring}
\label{methods3.7}

	The rewiring principle can be rigorously explained as follows. We first number each vertex in $G$ from 1 to |$G$| (where $n=|G|$, $n$ being the total number of vertices). If there is a connection between vertices $u$ and $v$ in $G$, we generate a random number between 0 and 1. If this random number is less than a given probability $p$, then the connection will be rewired.  Without loss of generality assume $u < v$.  We then fix the connection to vertex $u$, and move the connection to another vertex, $k$, such that $u$ and $k$ are now connected whereas they were not previously.

\subsection{Graph Energy}
\label{methods3.8}

	For a graph $G=(V,E$) where $V$ is the set of vertices and $E$ is the set of edges, one can define an arbitrary scalar function $u: V \rightarrow \mathbb{R}$. Consequently, one can then define the energy of $u$ associated with the graph.  The energy of $u$ associated with $G$, $\mathcal{E}(u)$  is defined as:
\[
\mathcal{E}(u) =\sum_{x,y \in E} (u(x)-u(y))^2. 
\]
We analyzed the energies of the Laplacian matrix eigenfunctions, thus $u=\varphi$.  

\subsection{Spacial Variance}
\label{methods3.9}
	
In order to discuss spacial variance, we must first define the resistance between two vertices on a graph.  Let $G = (V, E)$ be a graph and $x, y \in V$ . Then the resistance between $x$ and $y$, $d(x, y)$, is given by:
\[
d(x, y) = \mathcal{E}(h(x,y))^{-1}
\]
Where $h(x,y)$ is a harmonic function defined as follows:

Let $G = (V,E)$ be a graph and $x,y \in V$. Then the harmonic function corresponding to $(x, y)$ is a scalar function $h(x,y): V \rightarrow \mathbb{R}$ such that:   \newline
1. $h(x,y)(x) = 0$   \newline
2. $h(x,y)(y) = 1$   \newline
3. $\mathcal{E}(h)^{-1}, \text{ where } h \text{ is an arbitrary scalar function on } V , \text{ is maximized at } h(x,y)$.
\newline

Finding the harmonic function is equivalent to finding a vector $h$ such that $Lh=z$, where $z$ is a vector whose entries are all 0 except for those entries corresponding to $x$ and $y$.  This is analogous to what ``harmonic" means in Euclidean space.  This changes the maximization problem in condition 3. to solving a system of linear equations.  \newline

Using these we can now define the spacial variance of a graph. Again, let $G = (V,E)$ be a graph with $n$ vertices and $u$ be a scalar function on $V$. Let $\gamma$ be a constant. Then the $\gamma^{th}$ spacial variance of $u$ over $G$ is given by:
\[
Var_{\gamma}(u)= \frac{1}{n}\sum_{x,y \in E} d(x,y)^{\gamma} (u(x)-u(y))^2
\]
In this paper, the spacial variances of eigenfunctions of Laplacian matrices were evaluated at $\gamma=1$ and then analyzed.

%% file: CElegans.bbl
\begin{thebibliography}{10}
\providecommand{\url}[1]{\texttt{#1}}
\providecommand{\urlprefix}{URL }
\expandafter\ifx\csname urlstyle\endcsname\relax
  \providecommand{\doi}[1]{doi:\discretionary{}{}{}#1}\else
  \providecommand{\doi}{doi:\discretionary{}{}{}\begingroup
  \urlstyle{rm}\Url}\fi
\providecommand{\bibAnnoteFile}[1]{%
  \IfFileExists{#1}{\begin{quotation}\noindent\textsc{Key:} #1\\
  \textsc{Annotation:}\ \input{#1}\end{quotation}}{}}
\providecommand{\bibAnnote}[2]{%
  \begin{quotation}\noindent\textsc{Key:} #1\\
  \textsc{Annotation:}\ #2\end{quotation}}
\providecommand{\eprint}[2][]{\url{#2}}

\bibitem{MAND1}
Mandelbrot BB (1967) {{H}ow long is the coast of Britain? Statistical
  self-similarity and fractal dimension}.
\newblock Science 156: 636--638.
\bibAnnoteFile{MAND1}

\bibitem{LFKTW2}
Liebovitch LS, Fischbarg J, Koniarek JP, Todorova I, Wang M (1987) {{F}ractal
  model of ion-channel kinetics}.
\newblock Biochim Biophys Acta 896: 173--180.
\bibAnnoteFile{LFKTW2}

\bibitem{LLW3}
Lowen SB, Liebovitch LS, White JA (1999) {{F}ractal ion-channel behavior
  generates fractal firing patterns in neuronal models}.
\newblock Phys Rev E Stat Phys Plasmas Fluids Relat Interdiscip Topics 59:
  5970--5980.
\bibAnnoteFile{LLW3}

\bibitem{PU4}
Paramanathan P, Uthayakumar R (2008) {{A}pplication of fractal theory in
  analysis of human electroencephalographic signals}.
\newblock Comput Biol Med 38: 372--378.
\bibAnnoteFile{PU4}

\bibitem{BBG5}
Bernard F, Bossu JL, Gaillard S (2001) {{I}dentification of living
  oligodendrocyte developmental stages by fractal analysis of cell morphology}.
\newblock J Neurosci Res 65: 439--445.
\bibAnnoteFile{BBG5}

\bibitem{SLM6}
Smith TG, Lange GD, Marks WB (1996) {{F}ractal methods and results in cellular
  morphology--dimensions, lacunarity and multifractals}.
\newblock J Neurosci Methods 69: 123--136.
\bibAnnoteFile{SLM6}

\bibitem{FJ7}
Fernandez E, Jelinek HF (2001) {{U}se of fractal theory in neuroscience:
  methods, advantages, and potential problems}.
\newblock Instituto de Bioingenier\;ia 24: 309--321.
\bibAnnoteFile{FJ7}

\bibitem{KHA8}
Kiselev VG, Hahn KR, Auer DP (2003) {{I}s the brain cortex a fractal?}
\newblock Neuroimage 20: 1765--1774.
\bibAnnoteFile{KHA8}

\bibitem{FJ9}
Jelinek HF, Fern\'andez E (1998) {{N}eurons and fractals: how reliable and
  useful are calculations of fractal dimensions?}
\newblock J Neurosci Methods 81: 9--18.
\bibAnnoteFile{FJ9}

\bibitem{W10}
Werner G (2010) {{F}ractals in the nervous system: conceptual implications for
  theoretical neuroscience}.
\newblock Front Physiol 1: 15.
\bibAnnoteFile{W10}

\bibitem{M11}
Murray J (1995) Use and {A}buse of {F}ractal {T}heory in {N}euroscience.
\newblock The Journal of Comparative Neurology 361: 369--371.
\bibAnnoteFile{M11}

\bibitem{SB12}
Smith TG, Behar TN (1994) {{C}omparative fractal analysis of cultured glia
  derived from optic nerve and brain demonstrate different rates of
  morphological differentiation}.
\newblock Brain Res 634: 181--190.
\bibAnnoteFile{SB12}

\bibitem{SBLMS13}
Smith TG, Behar TN, Lange GD, Marks WB, Sheriff WH (1991) {{A} fractal analysis
  of cultured rat optic nerve glial growth and differentiation}.
\newblock Neuroscience 41: 159--166.
\bibAnnoteFile{SBLMS13}

\bibitem{RSSS14}
Reichenbach A, Siegel A, Senitz D, Smith TG (1992) {{A} comparative fractal
  analysis of various mammalian astroglial cell types}.
\newblock Neuroimage 1: 69--77.
\bibAnnoteFile{RSSS14}

\bibitem{CSEDHN15}
Caserta F, Stanley HE, Eldred WD, Daccord G, Hausman RE, et~al. (1990)
  {{P}hysical mechanisms underlying neurite outgrowth: {A} quantitative
  analysis of neuronal shape}.
\newblock Phys Rev Lett 64: 95--98.
\bibAnnoteFile{CSEDHN15}

\bibitem{B16}
Bieberich E (2002) {{R}ecurrent fractal neural networks: a strategy for the
  exchange of local and global information processing in the brain}.
\newblock BioSystems 66: 145--164.
\bibAnnoteFile{B16}

\bibitem{S17}
Sporns O (2006) {{S}mall-world connectivity, motif composition, and complexity
  of fractal neuronal connections}.
\newblock BioSystems 85: 55--64.
\bibAnnoteFile{S17}

\bibitem{FSCFS18}
Free SL, Sisodiya SM, Cook MJ, Fish DR, Shorvon SD (1996) {{T}hree-dimensional
  fractal analysis of the white matter surface from magnetic resonance images
  of the human brain}.
\newblock Cereb Cortex 6: 830--836.
\bibAnnoteFile{FSCFS18}

\bibitem{BO19}
Bullmore E, Sporns O (2009) {{C}omplex brain networks: graph theoretical
  analysis of structural and functional systems}.
\newblock Nat Rev Neurosci 10: 186--198.
\bibAnnoteFile{BO19}

\bibitem{SR20}
Stam CJ, Reijneveld JC (2007) {{G}raph theoretical analysis of complex networks
  in the brain}.
\newblock Nonlinear Biomed Phys 1: 3.
\bibAnnoteFile{SR20}

\bibitem{FCR21}
Fallani FV, Costa LF, Rodriguez FA, Astolfi L, Vecchiato G, et~al. (2010) {{A}
  graph-theoretical approach in brain functional networks. {P}ossible
  implications in {E}{E}{G} studies}.
\newblock Nonlinear Biomed Phys 4 Suppl 1: S8.
\bibAnnoteFile{FCR21}

\bibitem{SK22}
Sporns O, Kotter R (2004) {{M}otifs in brain networks}.
\newblock PLoS Biol 2: 1910-1918.
\bibAnnoteFile{SK22}

\bibitem{IA23}
Itzkovitz S, Alon U (2005) {{S}ubgraphs and network motifs in geometric
  networks}.
\newblock Phys Rev E Stat Nonlin Soft Matter Phys 71: 026117.
\bibAnnoteFile{IA23}

\bibitem{MOOFOK24}
Morita S, Oshio Ki, Osana Y, Funabashi Y, Oka K, et~al. (2001) {Geometrical
  structure of the neuronal network of \emph{Caenorhabditis elegans}}.
\newblock Physica A 298: 553--561.
\bibAnnoteFile{MOOFOK24}

\bibitem{VPC25}
{Varshney} LR, {Chen} BL, {Paniagua} E, {Hall} DH, {Chklovskii} DB (2011)
  {Structural Properties of the Caenorhabditis elegans Neuronal Network}.
\newblock PLoS Comput Bio 7 7: e1001066.
\bibAnnoteFile{VPC25}

\bibitem{DKC}
Das KC (2004) The {L}aplacian spectrum of a graph.
\newblock Comput Math Appl 48: 715--724.
\bibAnnoteFile{DKC}

\bibitem{ZD}
Zhou D (2008) Spectral analysis of {L}aplacians on certain fractals.
\newblock ProQuest LLC, Ann Arbor, MI, 109 pp.
\newblock
  \urlprefix\url{http://gateway.proquest.com/openurl?url_ver=Z39.88-2004&rft_val_fmt=info:ofi/fmt:kev:mtx:dissertation&res_dat=xri:pqdiss&rft_dat=xri:pqdiss:NR43396}.
\newblock Thesis (Ph.D.)--University of Waterloo (Canada).
\bibAnnoteFile{ZD}

\bibitem{BHS26}
Berry T, Heilman SM, Strichartz RS (2009) Outer approximation of the spectrum
  of a fractal {L}aplacian.
\newblock Experiment Math 18: 449--480.
\bibAnnoteFile{BHS26}

\bibitem{K27}
Koren Y (2005) Drawing graphs by eigenvectors: theory and practice.
\newblock Comput Math Appl 49: 1867--1888.
\bibAnnoteFile{K27}

\bibitem{M28}
Mohar B (1991) The {L}aplacian spectrum of graphs.
\newblock In: Graph theory, combinatorics, and applications. {V}ol.\ 2
  ({K}alamazoo, {MI}, 1988), New York: Wiley, Wiley-Intersci. Publ. pp.
  871--898.
\bibAnnoteFile{M28}

\bibitem{PST29}
Pisanki T, Shawe-Taylor J (2000) Characterizing {G}raph {D}rawing with
  {E}igenvectors.
\newblock J Chem Inf Comput Sci 40: 567--571.
\bibAnnoteFile{PST29}

\bibitem{PVC30}
Varshney L, Chen B, Paniagua E, Hall D, Chklovskii D (2011) Structural
  properties of the \emph{Caenorhabditis elegans} neuronal network.
\newblock \underline{http://mitedu/lrv/www/elegans/} .
\bibAnnoteFile{PVC30}

\bibitem{BMAB31}
Bassett DS, Meyer-Lindenberg A, Achard S, Duke T, Bullmore E (2006) {{A}daptive
  reconfiguration of fractal small-world human brain functional networks}.
\newblock Proc Natl Acad Sci USA 103: 19518--19523.
\bibAnnoteFile{BMAB31}

\bibitem{SH32}
Sporns O, Honey CJ (2006) {{S}mall worlds inside big brains}.
\newblock Proc Natl Acad Sci 103: 19219--19220.
\bibAnnoteFile{SH32}

\bibitem{WS33}
Watts DJ, Stogartz SH (1998) Collective dynamics of 'small-world' networks.
\newblock Nature 393: 440--442.
\bibAnnoteFile{WS33}

\bibitem{CS34}
{Cs{\'a}nyi} G, {Szendr{\H o}i} B (2004) {Fractal-small-world dichotomy in
  real-world networks}.
\newblock Phys Rev E Stat Nonlin Soft Matter Phys 70: 016122.
\bibAnnoteFile{CS34}

\bibitem{OSLT35}
Okoudjou KA, Saloff-Coste L, Teplyaev A (2008) Weak uncertainty principle for
  fractals, graphs and metric measure spaces.
\newblock Trans Amer Math Soc 360: 3857--3873.
\bibAnnoteFile{OSLT35}

\bibitem{BKNPPT}
{Begue} M, {Kelleher} DJ, {Nelson} A, {Panzo} H, {Pellico} R, et~al. (2011)
  {Random walks on barycentric subdivisions and the Strichartz hexacarpet}.
\newblock ArXiv e-prints .
\bibAnnoteFile{BKNPPT}

\end{thebibliography}
